\documentclass{article}

\usepackage{arxiv}

\usepackage[utf8]{inputenc} 
\usepackage[T1]{fontenc}    
\usepackage{hyperref}       
\usepackage{url}            
\usepackage{booktabs}       
\usepackage{amsfonts}       
\usepackage{nicefrac}       
\usepackage{microtype}      
\usepackage{lipsum}		
\usepackage{graphicx}
\usepackage{natbib}
\usepackage{doi}
\usepackage{xcolor}
\usepackage[normalem]{ulem}
\usepackage{amsmath}

\title{Leveraging language representation for material recommendation, ranking, and exploration}
 
\author{ \href{https://orcid.org/0000-0002-3304-4996}{\includegraphics[scale=0.1]{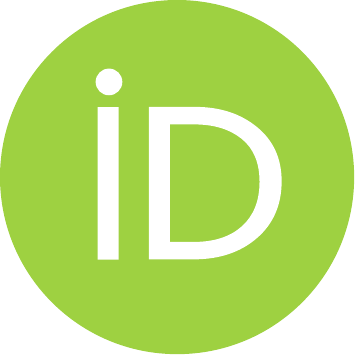}\hspace{1mm}Jiaxing Qu$^{1}$}\thanks{These authors share equal contributions to this work.} \\
	\And
	\href{https://orcid.org/0000-0003-1664-9114}{\includegraphics[scale=0.1]{orcid.pdf}\hspace{1mm}Yuxuan Richard Xie$^{2}$\footnotemark[1]}\\
 	\And
    \href{https://orcid.org/0000-0002-9787-5967}
    {\includegraphics[scale=0.1]{orcid.pdf}\hspace{1mm}Kamil M. Ciesielski$^{4}$}\\
 	\And
    \href{https://orcid.org/0000-0002-7193-579X}
    {\includegraphics[scale=0.1]{orcid.pdf}\hspace{1mm}Claire E. Porter$^{4}$}\\
 	\And
    \href{https://orcid.org/0000-0003-0826-2446}
    {\includegraphics[scale=0.1]{orcid.pdf}\hspace{1mm}Eric S. Toberer$^{4}$}\\
 	\And
	\href{https://orcid.org/0000-0002-7816-1803}{\includegraphics[scale=0.1]{orcid.pdf}\hspace{1mm}Elif Ertekin$^{1, 3}$} \\
}

\renewcommand{\headeright}{}
\renewcommand{\shorttitle}{}

\hypersetup{
pdftitle={Leveraging Language representations for material recommendation and exploration: a case study in thermoelectric materials},
pdfsubject={q-bio.NC, q-bio.QM},
pdfauthor={},
pdfkeywords={First keyword, Second keyword, More},
}

\begin{document}
\maketitle
\let\thefootnote\relax\footnotetext{$^1$ Department of Mechanical Science and Engineering, $^2$ Beckman Institute for Advanced Science and Technology, $^3$ Materials Research Laboratory, University of Illinois Urbana-Champaign, Urbana, IL, 61802, $^4$ Colorado School of Mines, Golden, CO, 80401}

\begin{abstract}
Data-driven approaches for material discovery and design have been accelerated by emerging efforts in machine learning. 
However, general representations of crystals to explore the vast material search space remain limited. We introduce a material discovery framework that uses natural language embeddings derived from
language models as representations of compositional and structural features. 
The discovery framework consists of a joint scheme that
first recalls relevant candidates, and next ranks the candidates based on multiple target properties.
The contextual knowledge encoded in language representations
conveys information about material properties and structures, enabling both representational similarity analysis for recall, and multi-task learning to share information across related properties. 
By applying the framework to thermoelectrics, we demonstrate diversified recommendations of prototype structures and identify under-studied high-performance material spaces.
The recommended materials
are corroborated by first-principles calculations and experiments, revealing novel materials with potential high performance.
Our framework provides a task-agnostic means for effective material recommendation and can be applied to various material systems. 
\end{abstract}




\section{Introduction}


Rapid growth of data in materials science has opened a data-centric paradigm (\cite{hey2009fourth}) for discovery of novel materials.
In this paradigm, machine learning (ML) models trained on large material data sets can computationally screen candidates for field-specific applications such as batteries (\cite{aykol2020machine}), thermoelectrics,  (\cite{wang2020machine}) and solar cells (\cite{mahmood2021machine}), etc. 
The key objective of the model-driven approach is to identify candidates that exhibit targeted, desirable material properties. 
Extracting representative features of materials to capture attributes is therefore a key to success of accurate model performance and property prediction. 
Conventionally, material feature extraction has consisted of hand-crafted descriptors that contain essential information related to composition and crystal structure, relying on physical and mathematical intuition (\cite{schmidt2019recent, behler2007generalized, isayev2017universal}).
Until recently, materials' atomic structures have been treated as graphs, where convolution operations extract features from local chemical environments for accurate property predictions (\cite{xie2018crystal}).
Subsequently, several models have been proposed to directly learn features from material compositions or structures for supervised prediction tasks (\cite{chen2019graph,choudhary2021atomistic,yan2022periodic,jha2018elemnet}). 
An outstanding challenge, however, is to identify a universal and task-agnostic representation that can enables generalized efficient search and navigation of the vast and largely unlabeled material space.


Advances in natural language processing have allowed information mining from the large corpus of material science related literature in an unsupervised fashion. 
A pioneering work utilizes word embeddings trained on a large material text corpus to encode material science knowledge into information-dense vector representations (\cite{tshitoyan2019unsupervised}). 
Given a context word for technological application, e.g. ``thermoelectrics'', candidate materials are ranked by similarity to the word embedding of the context word. 
Word embeddings obtained on material compositions have also shown competitive performance on material property prediction tasks (\cite{wang2021compositionally}). 
However, word embedding, such as in Word2Vec (\cite{tshitoyan2019unsupervised}), does not capture the contextual meaning of the word that is present in a sentence.
Progress on contextual embedding models has been enabled by masked language modeling to train Transformer-based language models (MatBERT -- \cite{trewartha2022quantifying}, MatSciBERT -- \cite{gupta2022matscibert}) for material discovery and knowledge extraction from millions of unstructured material science literatures. 
By employing pretrained BERT models, latent knowledge learnt from the material science text corpus can be encoded into the representation and then subjected to a number of subsequent prediction tasks. 



\begin{figure}[!t]
	\centering
        \includegraphics[width=\columnwidth]{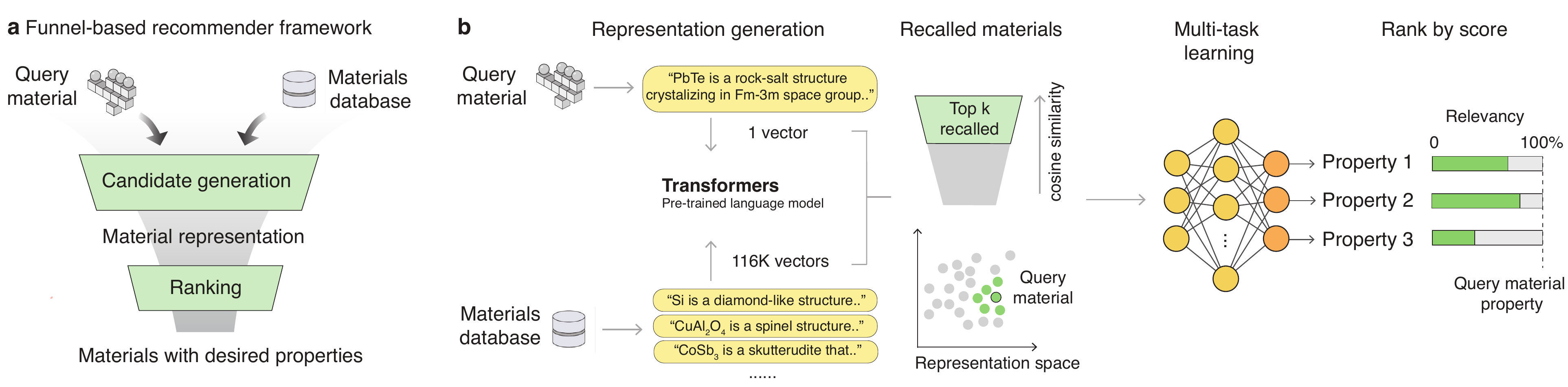}
	\caption{\textbf{a}, The proposed funnel-based recommender framework in which candidate materials are recalled, and ranked based on 
    similarity to the query material. \textbf{b}, The schematic workflow to screen candidate materials including constructing language representations, recalling candidates, and multi-task prediction for ranking.}
	\label{fig:fig1_overview}
\end{figure}

In the context of new high-performance materials, some essential factors should be included in the recommendation pipeline: (i) effective representations of both chemical and structural complexity in the large material space, (ii) successful recall of relevant candidates to the query material or property of interest,  
and (iii) accurate candidate ranking based on multiple desired functional properties. 
Previously, recommender-like systems for materials were developed 
to filter by identifying materials for which predicted confidence levels of target properties fall within a desirable range for thermoelectrics (\cite{gaultois2016perspective}), to predict chemically relevant compositions for pseudo-ternary systems (\cite{seko2018compositional, seko2018matrix}), and to propose experimental synthesis conditions (\cite{hayashi2019recommender}). 
However, a systematic and generalizable recommendation approach, which incorporates the factors mentioned above for representation, recall, and ranking, could accelerate discovery of desirable material candidates across diverse applications.
Here we present a material recommendation framework that leverages language representations of composition and structure to explore a large space and identify similar candidate materials, given a query material with targeted   desired properties.  
The framework invokes a funnel-based architecture comprising a candidate generation (``recall'') step and a subsequent property evaluation (``ranking'') step (Figure \ref{fig:fig1_overview}a). 
We first constructed representations for $\sim$116,000
materials using text description as the input to the transformer based language models. 
By evaluating different embedding methods on various downstream tasks, we found that material language representations are both highly potent in recalling relevant material candidates, and capable of predicting properties with comparable performance to state-of-the-art specialized ML models. 
For improved ranking, we introduced a multi-gate mixture-of-experts (MMoE) model, a multi-task learning strategy, to exploit correlations between material property prediction tasks (Figure \ref{fig:fig1_overview}b). 
As a demonstration example of material discovery, we applied our framework to search and recommend high-performance thermoelectrics (TEs) -- materials that convert waste heat into electricity. 
Using this framework, we successfully identify structurally-diversified TE candidates that are relevant to query materials. 
Additionally, we identify and further explore several under-searched high-performance materials spaces including halide perovskite, delafossite-like, and spinel-like structures as promising TE candidates. 
As an evaluation of the  effectiveness of this framework, we performed first-principles calculations and experiments on the recommended materials and successfully identified CuZn$_2$GaTe$_4$ as a new TE material that, under further optimization, may demonstrate high performace.

\begin{figure}[!b]
	\centering
        \includegraphics[width=0.85\columnwidth]{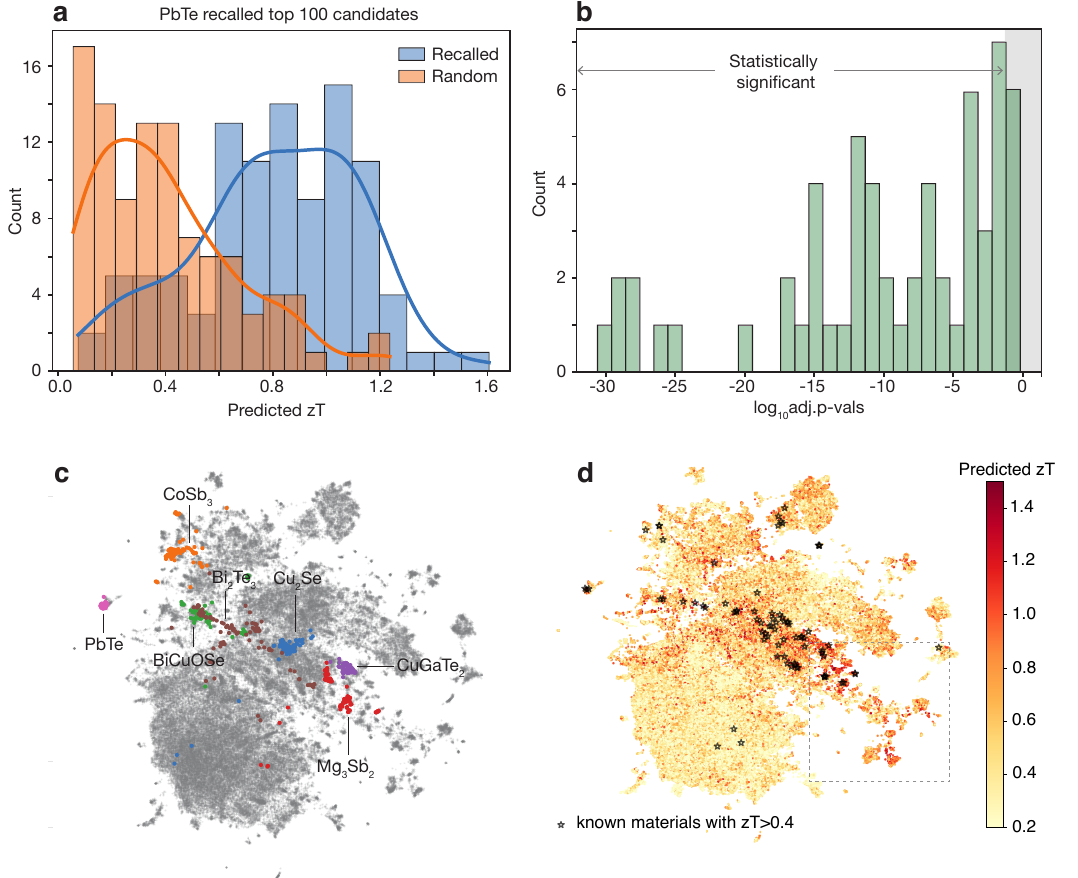}
	\caption{\textbf{a}, Distributions of predicted $zT$ of the top-100 recalled candidates for PbTe as the query material and randomly sampled 100 materials. The distributions are also visualized in kernel density estimation. Predicted $zT$ are obtained from the MMoE models (section \ref{result:mmoe}). \textbf{b}, Adjusted $p$-values of the candidates for the top-100 highest $zT$ materials from experimental UCSB and ESTM datasets. Out of 100 materials, 94 show significantly different $zT$ distributions from random. \textbf{c}, Recall results of seven high-performing TE materials are highlighted on the UMAP projection of 116K material representations. Each color corresponds to first 100 materials recalled via cosine similarity. The UMAP projection is obtained on the best performing embeddings (MatBERT, structure embedding) from Section \ref{results:embeddings}. \textbf{d}, UMAP overlaid with the predicted $zT$. All known materials with $zT$>0.4 from the experimental datasets are indicated by stars.}
	\label{fig:fig6_stats}
\end{figure}


\section{Results and discussions}
\label{sec:others}

\subsection{A language-based framework enables material recommendations and discovery.}


Machine learning-based recommender systems leverage a large corpus of training data to provide precise suggestions when querying for items among a large candidate pool  (\cite{covington2016deep, gomez2015netflix}).  
During the recommendation process, a funnel-based architecture is typically applied for initial screening, followed by more fine-grained ranking steps. 
Inspired by the standard design of recommender systems, we adapted our framework to material science to effectively search a large space and recommend relevant materials with similar functional performance to a query material. 
Specifically, we designed a funnel-based architecture that can be decoupled into a recall step and a ranking step (Figure \ref{fig:fig1_overview}a). 
To enable candidate recall for a query material, we converted each material into text-based descriptions that include both compositional and structural information. 
Using language models (\cite{trewartha2022quantifying, gupta2022matscibert}) pretrained on material science literature, we then obtained output embeddings on these text-based material descriptions. 
These embeddings encode contextual representations to capture compositional and structural features with high-level interactions arising from self-attention (\cite{devlin2018bert}). 
In the recall step (``candidate generation''), candidates can be searched via cosine similarity against the query material in the representation space (Figure \ref{fig:fig1_overview}b).
In the ranking step, recalled candidate materials are evaluated and ranked using a multi-objective scoring function trained on the encoded representations to simultaneously predict multiple material properties through neural networks. 
For this work, we exploited task correlations between predicting five TE properties by training multi-task learning MMoE models, which provided improved accuracy compared to models trained on single tasks.


\par To obtain compositional and structural level representations for the database consisting of 116K materials (Section \ref{method:dataset}), we embedded all material formulae (e.g., "PbTe") and sentence descriptors automatically generated (Robocrystallographer -- \cite{ganose2019robocrystallographer})
from the structures (e.g., "PbTe is Halite, Rock Salt structured and crystallizes in the cubic $Fm\bar{3}m$ space group...") as the input to pretrained language models. 
Embedding each formula or structure generates a dense vector output from the model hidden layer, which contains latent material-specific knowledge learnt during unsupervised pretraining. 
In Figure \ref{fig:fig6_stats}, we demonstrate that recalled candidates in the representation space are not only compositionally and structurally related to the query material, but also can exhibit similar functional performance to a query material. 
Starting with known materials with favorable properties for TEs such as PbTe, we analyzed the top recalled candidates and found significantly different predicted figure-of-merit $zT$ distributions from random sampling as indicated by $p$-values (Figure \ref{fig:fig6_stats}a). 
We repeated this experiment for a total of 100 materials with the known highest $zT$; 94 of these show statistical significance with $p$<0.05 (Figure \ref{fig:fig6_stats}b), showing that recalled materials show distributions that are distinct from random.
Moreover, low-dimensional Uniform Manifold Approximation and Projection (UMAP) (\cite{mcinnes2018umap}) of the material representations display latent signatures of seven high-performing TE materials along with their top-100 recalled materials, each indicated by a different color (Figure \ref{fig:fig6_stats}c). 
We further observed a distinct clustering pattern, in which known materials with good $zT$ (>0.4) form a ``band'' in the projection (Figure \ref{fig:fig6_stats}d). 
Additionally, the observed ``band'' overlaps with the MMoE-predicted high $zT$ (also Figure \ref{fig:fig6_stats}d). 
The distribution of $zT$ in the representation space provides opportunities to explore under-explored material spaces, such as the region enclosed in the grey box with high predicted $zT$. 

To understand how individual steps contribute to the performance of our material recommendation framework, in the following we assess the effectiveness of different representation strategies, recall ability, and property prediction via multi-task learning. 
Further, we demonstrate our framework to search, ranking, and exploration tasks for TE materials.




\subsection{Language models offer effective representations of material composition, structure, and properties.}\label{results:embeddings}



\begin{figure}[!b]
	\centering
        \includegraphics[width=\columnwidth]{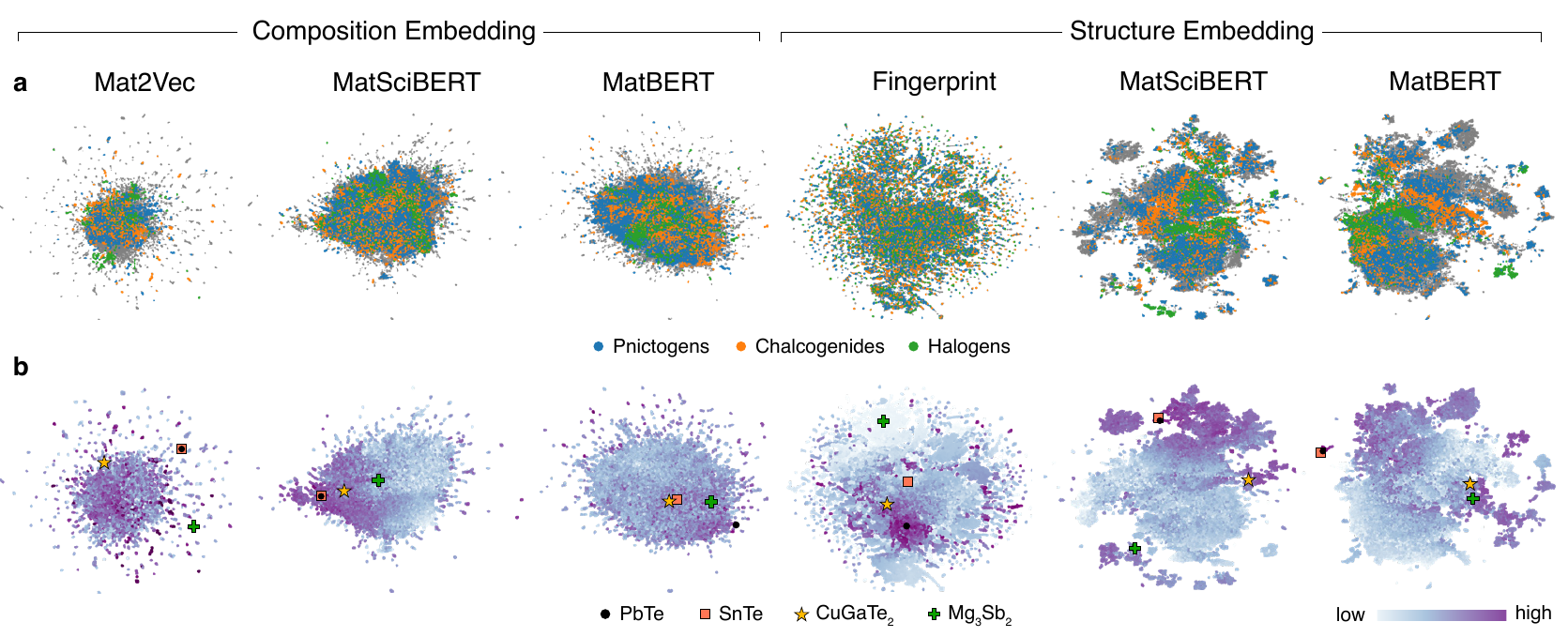}
	\caption{UMAP projections of 116K materials using different embedding models. 
 Materials are colored   \textbf{a}, by anionic groups, and \textbf{b}, by cosine similarity to PbTe under the current embedding model.}
	\label{fig:fig2_umap_embedding}
\end{figure}


Effective representations require rendering information about material design principles and intrinsic properties. 
We evaluated several strategies for material representation, focusing on unsupervised generation of features to convey diverse chemical and structural information.  
In total, we investigated six embedding methods. 
For composition level representation, we embed the material formula using pretrained word embedding Mat2Vec (\cite{tshitoyan2019unsupervised}) and contextualized word embedding from MatSciBERT (\cite{gupta2022matscibert}) and MatBERT (\cite{trewartha2022quantifying}). 
For structural level representation, we obtained local environment based structure fingerprints (\cite{zimmermann2020local}) and sentence embeddings of text-based material descriptions from MatSciBERT and MatBERT. 
Note that for BERT models, we constructed embedding vectors from entire passages of text consisting of human-readable crystal structure characteristics (\cite{ganose2019robocrystallographer}), as described in Section \ref{method:pretrained_languagemodels}.

To assess whether the embedding models have encoded material knowledge in the representations, we projected the six different material embedding vectors into low-dimensional spaces with UMAP, as visualized in Figure \ref{fig:fig2_umap_embedding}.
Embedded materials consisting of groups 15 (pnictogen), 16 (chalcogen), and 17 (halogen) on the periodic table are indicated by color (Figure 
 \ref{fig:fig2_umap_embedding}a). 
Overall, structure level representations exhibit more distinct separation (well-defined domains) by material groups, apart from  fingerprints which are solely determined by structural similarity and include information only about local but not semi-local and global environments. 
By contrast, composition level representations retain the expected chemical differences, but form more disperse and heterogeneous clusters.

To better interpret the embedding results, we picked three well-studied TE materials, including SnTe -- a rock-salt structural analog of PbTe with highest reported $zT$ of $\sim$1.8 (\cite{tang2018manipulation}), CuGaTe$_2$ -- a diamond-like semiconductor in chalcopyrite structure that achieves a $zT$ of 1.5 (\cite{wu2022significantly}), as well as Mg$_3$Sb$_2$ -- a layered Zintl phase with the highest $zT$ of $\sim$1.65 (\cite{zhang2017discovery, ohno2018phase}), and visualized their proximity to PbTe in the representation space (Figure 
 \ref{fig:fig2_umap_embedding}b). 
All three materials have demonstrate high $zT$ around 1.5, but the high performance arises from different combinations of properties relevant to TEs (i.e. electronic and thermal transport) due to their different structures. 
Embedded materials in the representation space follow our anticipated similarity (PbTe$\approx$SnTe > CuGaTe$_2$ > Mg$_3$Sb$_2$), apart from fingerprints and composition embeddings from MatBERT. 

\par Next, we quantitatively evaluated material embedding performance on downstream property prediction tasks. We applied a feature-based-approach to train regression models directly on the derived embeddings, instead of optimizing BERT parameters on the task-specific loss, i.e., fine-tuning (\cite{devlin2018bert}). 
This approach is more computationally efficient due to fixed features, and grants flexibility to adapt task-specific architectures or combine features of various sources across different models.
We list the cross validation performance on predicting six material properties for 5,700 materials in Table \ref{tab:table_benchmark}. 
The task models were multi-layer perceptrons (MLPs) with mean-absolute-error (MAE) training loss. 
The tasks consisted of 
band gap, energy per atom, bulk modulus, shear modulus, Debye temperature, and coefficient of thermal expansion from AFLOW dataset (\cite{curtarolo2012aflow}). 
Performance metrics of models trained using several embeddings, such as structure embeddings extracted from MatBERT, achieved accurate performance.
Moreover, by leveraging latent material science knowledge embeddings from pretrained large language models, the language representation supports learning in the face of data scarcity, a ubiquitous challenge in applying ML to materials science. 
For small data with only 200 training materials, models trained using these embeddings outperform graph neural networks (CGCNN -- \cite{xie2018crystal}) when tested on 100 independent materials (Supplementary Figure S1). 
These results suggest that pretrained language models, in combination with text-based structure descriptions, provide a competitive avenue to generate features for material representations. 



\renewcommand{\arraystretch}{1.3} 
\begin{table}
	\caption{Benchmarking six different embedding models on six regression property prediction tasks with MAEs and R$^2$-scores. ($E$/atom -- energy per atom (eV), $E_\mathrm{g}$ -- band gap (eV), $K$ -- bulk modulus (GPa), $G$ -- shear modulus (GPa), $\theta$ -- Debye temperature (K), $\alpha$ -- coefficient of thermal expansion (K$^{-1}$)) }
	\centering
	\begin{tabular}{p{1.3cm} p{1.2cm} p{1.8cm} p{1.8cm} p{1.8cm} p{1.8cm} p{1.8cm}p{1.8cm}}
		\toprule
            \multicolumn{2}{c}{} & \multicolumn{3}{c}{Composition embedding} &
              \multicolumn{3}{c}{Structure embedding} \\
		Property   &   Metric &   Mat2Vec    &   MatSciBERT &   MatBERT   & Fingerprint   &   MatSciBERT  & MatBERT   \\
		\midrule
		$E$/atom  &   MAE &   0.47$\pm$0.02   &  0.42$\pm$0.01  &   0.37$\pm$0.01 &   1.13$\pm$0.02   &    0.32$\pm$0.02    &   \textbf{0.29$\pm$0.03 }   \\
  		&   R$^2$   & 0.81$\pm$0.02  &  0.86$\pm$0.01  &    0.88$\pm$0.01 &   0.283$\pm$0.02    &    0.95$\pm$0.01 &  \textbf{0.96$\pm$0.01 }\\
		$E_\mathrm{g}$    &   MAE    &   \textbf{0.15$\pm$0.01}    &  0.20$\pm$0.02  &   0.19$\pm$0.01   &    0.54$\pm$0.03   &  0.25$\pm$0.01   &   0.23$\pm$0.01   \\
  		&   R$^2$   &  \textbf{0.92$\pm$0.02} &  0.88$\pm$0.02  &    0.88$\pm$0.01    &    0.45$\pm$0.04    &    0.88$\pm$0.01 &   0.89$\pm$0.01    \\
		log\_$K$   &   MAE    &   0.18$\pm$0.01 &  0.18$\pm$0.01  &   0.17$\pm$0.01 &   0.45$\pm$0.01    &    0.16$\pm$0.01     &\textbf{0.15$\pm$0.01}   \\
  		&   R$^2$   & 0.83$\pm$0.01  &  0.83$\pm$0.03  &    0.85$\pm$0.02    &    0.26$\pm$0.02    &    0.90$\pm$0.01 &  \textbf{0.93$\pm$0.01}    \\
		log\_$G$   &   MAE   &   \textbf{0.20$\pm$0.01}    &  0.23$\pm$0.01 &   0.22$\pm$0.01    &    0.48$\pm$0.01    &   0.24$\pm$0.01 &   0.23$\pm$0.01    \\  
  		&   R$^2$   &  0.82$\pm$0.01 &  0.80$\pm$0.01  &    0.81$\pm$0.02    &    0.29$\pm$0.03    &    0.83$\pm$0.01 &   \textbf{0.84$\pm$0.01}    \\
		log$_\mathrm{10}$\_$\theta$  &   MAE  & 0.06$\pm$0.01  &  0.07$\pm$0.01  &   0.06$\pm$0.01    &    0.13$\pm$0.01  &  0.07$\pm$0.01   &   \textbf{0.06$\pm$0.01}    \\    
  		&   R$^2$   & 0.81$\pm$0.02  &  0.82$\pm$0.03  &    0.84$\pm$0.02    &    0.34$\pm$0.05    &    0.85$\pm$0.03 &   \textbf{0.88$\pm$0.02 }   \\
    	log$_\mathrm{10}$\_$\alpha$   &   MAE   & 0.07$\pm$0.01  &  0.07$\pm$0.01 &   0.07$\pm$0.01    &    0.15$\pm$0.01  &  0.07$\pm$0.01    &   \textbf{0.06$\pm$0.01}    \\    
  		&   R$^2$   &  0.78$\pm$0.03 &  0.81$\pm$0.02  &    0.81$\pm$0.02    &    0.19$\pm$0.02    &    0.87$\pm$0.03 &   \textbf{0.90$\pm$0.01}    \\
		\bottomrule
	\end{tabular}
	\label{tab:table_benchmark}
\end{table}


\subsection{Unsupervised candidate recall extracts highly relevant materials.}\label{results:evaluation}


For the recall (candidate generation) step, we use an unsupervised approach, as the models need to generalize to unseen query materials without further training while correctly recalling relevant candidates. 
In the unsupervised context, each query material is an individual prediction task, where the goal is to find a set of related materials.
While material recall is strictly based on cosine similarity in embedding space from text-based descriptions of composition and/or structure, we hypothesize that these embeddings contain latent material science knowledge, in which recalled materials will also share some similarities in properties to the query material. 



For commercial recommender systems, online learning (\cite{xiao2018personalized, song2014online}) makes data collection and model evaluation straightforward.
For our framework, we evaluate recalled TE materials in an offline setting with predefined `relevancy' (Section \ref{method:evaluation}) as a measure of the the composite differences in TE properties. 
We considered five TE properties -- power factor, Seebeck coefficient, electrical conductivity, thermal conductivity, and $zT$ for 826 unique host materials (Section \ref{method:dataset}) 
For each query material, the relevancy is obtained as the summation of absolute differences of these five properties, \textit{i.e.} candidates with similarity across all properties are considered most relevant. 

\begin{figure}
	\centering
        \includegraphics[width=0.75\columnwidth]{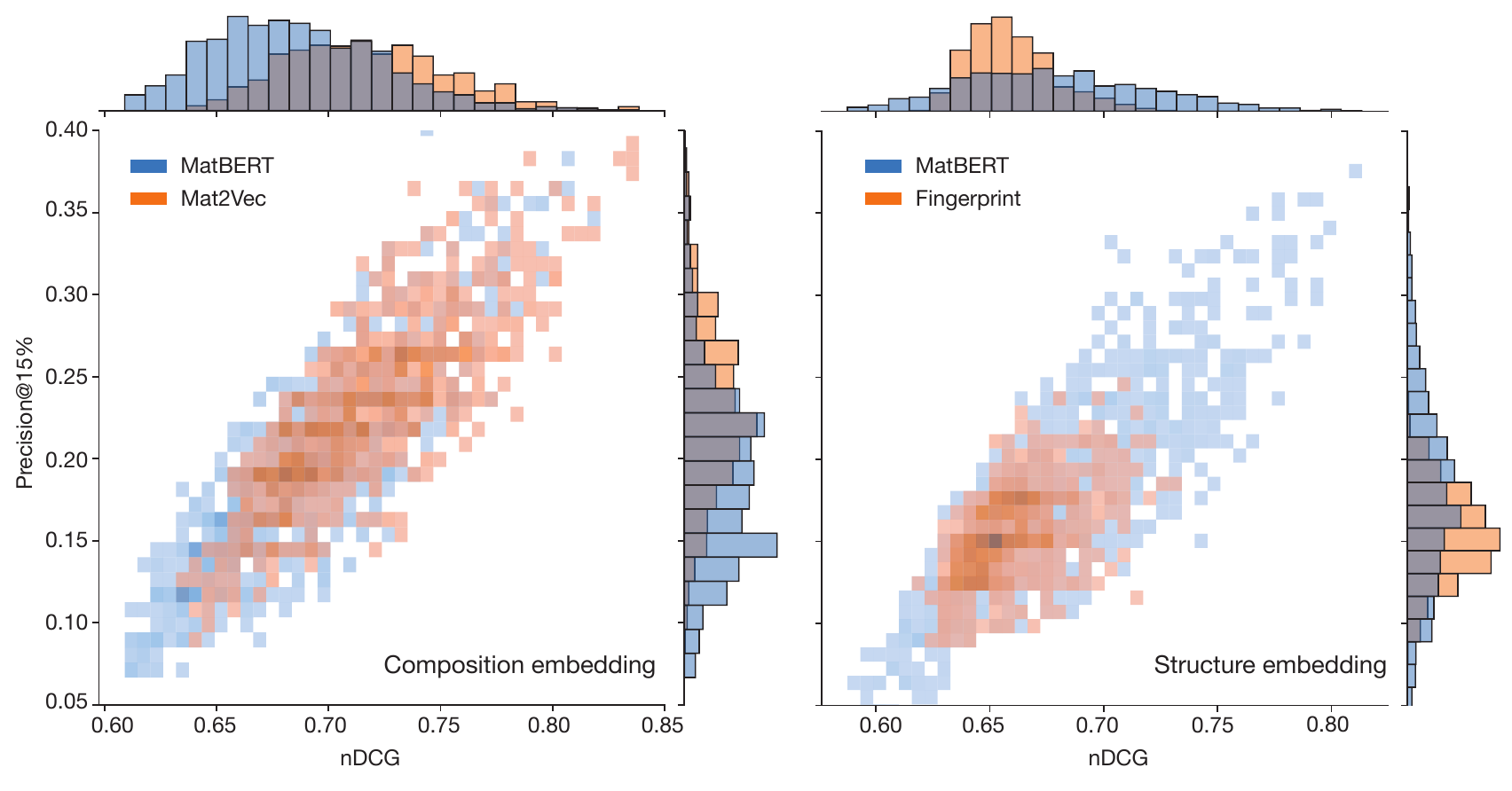}
	\caption{Evaluation metric of candidate generation stage for 116K materials. Each square represents a different query material. The X and Y-axis represent precision at 15\% and nDCG, which measure the accuracy and relative ranking order respectively. Evaluations are performed for composition embedding and structure embedding.}
	\label{fig:fig3_recall_metric}
\end{figure}

For evaluation, precision and normalized discounted cumulative gain (nDCG) were used as recall performance metrics (see Section \ref{method:evaluation}). 
Specifically, we calculated precision$@$15\% to assess the recall accuracy by defining the top 15\% of 826 materials (124 materials) as `relevant' to the query material based on experimental TE properties. 
We then evaluated the overlap between the top 124 recalled and relevant materials. 
A higher precision$@$15\% score indicates that more relevant materials have been recalled.
On the other hand, nDCG evaluates the ranking from the perspective of relative positions of items in the similarity-based list.
The two metrics are jointly visualized in Figure \ref{fig:fig3_recall_metric} and analyzed separately for composition and structure embeddings. 
Each scatter point denotes performance for one queried material. 
Ideally, candidates should have high precision$@$15\% and nDCG (top right corner). 
Using composition embeddings, Mat2Vec and MatBERT gave similar performance on both precision$@$15\% and nDCG, indicating the effectiveness of Mat2Vec word embedding in capturing latent material science knowledge. 
For structure embeddings, however, MatBERT recalls significantly more relevant materials than using fingerprint. 
This performance is nor surprising, since fingerprints only contain information about local structure at the motif level, but lack information at the semi-local level (i.e. motif connectivity) and global level (e.g. space group). 
From both composition and structure MatBERT embeddings, a considerable number of materials achieved precision$>$0.25 and nDCG$>$0.7, suggesting that the representations extracted similarity preserving signals which could be utilized for unsupervised search for similarly performing materials.


\subsection{Multi-task learning exploits cross-task correlations for improved property predictions.}\label{result:mmoe}



For a more accurate candidate material ranking, in the second stage of the funnel approach of Figure \ref{fig:fig1_overview} we improved multi-property predictions through multi-task learning.
Learning from multiple related tasks provides superior performance over single-task learning by modeling task-specific objectives and cross-task relationships (\cite{ma2018modeling, caruana1997multitask}). 
Multi-task learning is thus ideal to learn the underlying commonalities across different yet correlated material properties, improving performance for each task. 
\cite{sanyal2018mt} showed that joint-training on
several material properties leads to better model performance in prediction tasks. A mixture-of-experts framework (\cite{chang2022towards}) demonstrates transferability between models trained on different material properties, thereby improving task performance.

\begin{figure}
	\centering
        \includegraphics[width=\columnwidth]{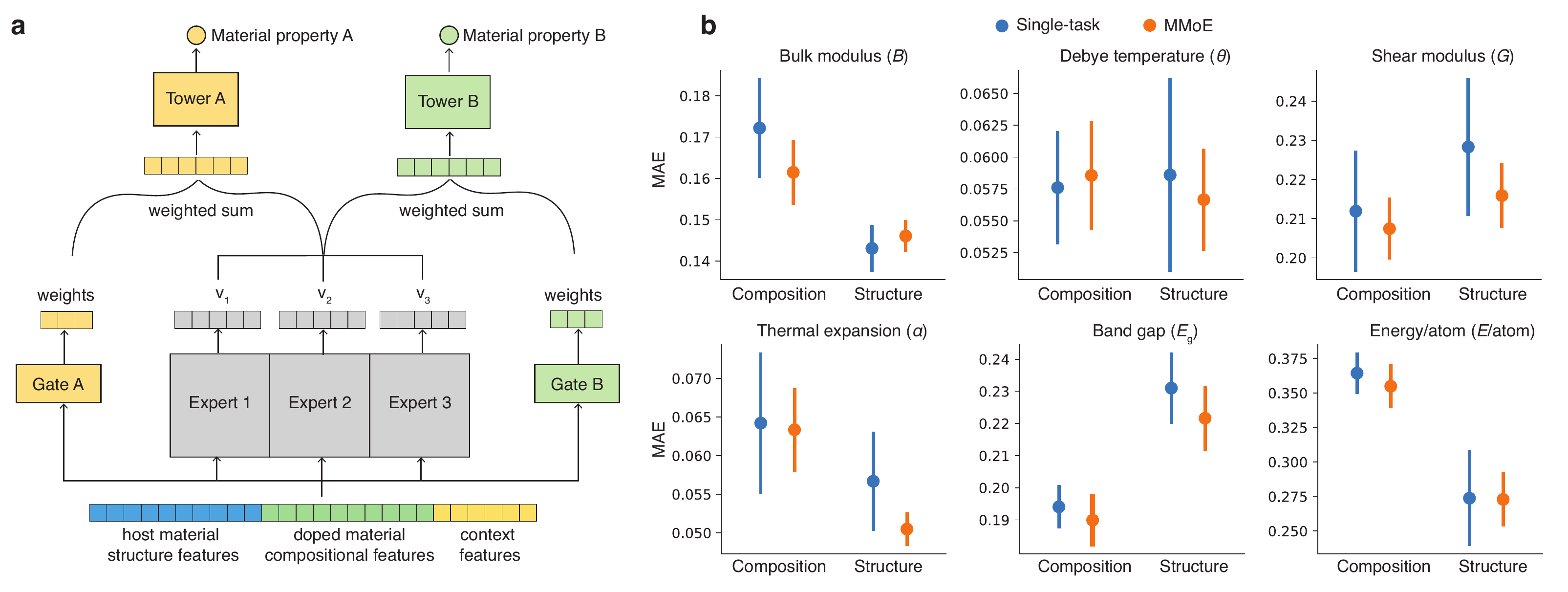}
	\caption{\textbf{a}, Schematic of the MMoE model. The gating networks learn contributions from different experts through the task-specific tower networks. \textbf{b}, Comparison of model performance for 6 material properties prediction between single-task models and MMoE using composition or structure embeddings.}
	\label{fig:fig4_mmoe_schematic_comparison}
\end{figure}

To this aim, we introduce multi-task learning with the MMoE model, which contains a set of expert networks and gating networks (Figure \ref{fig:fig4_mmoe_schematic_comparison}a). 
Through task-specific tower networks, the gating network for each property prediction allows the model to learn mixture contributions from different experts, thus exploiting the interconnections between tasks (Section \ref{method:mmoe}). 
In the approach adopted here, the input representations for MMoE models, discussed later in this section, are concatenated composition and structure embeddings, as well as context features for growth conditions (Figure \ref{fig:fig4_mmoe_schematic_comparison}a). 
We first benchmarked MMoE with single-task prediction to predict the six properties shown in Table \ref{tab:table_benchmark}.
As shown in Figure \ref{fig:fig4_mmoe_schematic_comparison}b, the MMoE results are within error of the single-task results, but show modest improvement by around 5-10\% for most cases.  
MMoE does show notably better model stability, indicated by lower variance in cross-validation performance. 
The complete single-task and MMoE performance can be found in Supplementary Figure S2, S3.


Next, we purposed MMoE for multi-task learning of thermoelectric properties.
The efficiency of TE energy conversion is given by figure of merit ($zT$) as: $zT$=$S^2\sigma T$/$\kappa$, where $S$ is Seebeck coefficient, $\sigma$ is electrical conductivity, $\kappa$ is thermal conductivity, and $T$ is the temperature.
A high $zT$ indicates a good thermoelectric, however, the properties that lead to high $zT$ are inter-dependent and often conflicting (\cite{snyder2008complex}).
For example, thermal conductivity increases with electrical conductivity as carrier concentrations approach the degenerate regime. 
Optimizing for TE performance is thus a challenging task that requires a balance of several properties. 
For this reason, we speculate that multi-task learning can naturally leverage the TE task correlations for better model performance.
We found moderate Pearson correlation ranging from 0.15 to 0.5 between the five TE properties considered here (Supplementary Figure S4), which is considered ideal for multi-task learning.
Interestingly, we found that multi-task learning significantly enhances the  predictive performance of Seebeck coefficient by 71\% compared with single-task prediction, with close performance for the other four tasks within variance from cross-validation (Supplementary Figure S5). 


\renewcommand{\arraystretch}{1.3} 
\begin{table}[!t]
	\caption{MAEs and R$^2$-scores of four representation methods as the input for MatBERT model to predict the thermoelectric properties on the UCSB and ESTM dataset. Note that for all representations, context features for temperature  are also included. (Power factor -- PF (S$^2$/m$^2$), Seebeck coefficient -- $S$ ($\mu$V/K), Electrical conductivity -- $\sigma$ (S/m), Thermal conductivity -- $\kappa$ (W/mK), figure-of-merit -- $zT$)}
	\centering
	\begin{tabular}{p{1.4cm} p{1.4cm} p{2.7cm} p{2.7cm} p{2.7cm} p{2.8cm}}
		\toprule
		Property   &   Evaluation   &   Host    &   Host    &   Doped  &   Host structure$+$\\
            &   &   composition  &   structure  &   composition &   Doped composition\\
		\midrule
		log\_PF   &   MAE  &   0.580$\pm$0.035    & 0.566$\pm$0.045 &   0.471$\pm$0.039    &    \textbf{0.433$\pm$0.024}    \\
  		&   R$^2$  &    0.584$\pm$0.080 &   0.624$\pm$0.095 &   0.740$\pm$0.060    &    \textbf{0.778$\pm$0.063} \\
		$S$   &   MAE &   52.3$\pm$6.1    & 53.3$\pm$8.0   &    36.8$\pm$5.1   &  \textbf{35.4$\pm$3.4}   \\
  		&   R$^2$  &    0.741$\pm$0.069 &   0.753$\pm$0.032    &    0.862$\pm$0.070    &    \textbf{0.872$\pm$0.046}    \\
		log\_$\sigma$   &  MAE &   1.151$\pm$0.076    &   1.157$\pm$0.063 &   0.693$\pm$0.040    &    \textbf{0.654$\pm$0.074} \\
  		&   R$^2$  &    0.576$\pm$0.080 &   0.585$\pm$0.076    &    0.813$\pm$0.036    &    \textbf{0.832$\pm$0.044}    \\
		log\_$\kappa$  &   MAE &   0.270$\pm$0.020    &   0.272$\pm$0.029    &    0.237$\pm$0.014    &    \textbf{0.221$\pm$0.022}    \\  
  		&   R$^2$  &    0.779$\pm$0.051 &   0.772$\pm$0.049    &    0.824$\pm$0.018    &    \textbf{0.841$\pm$0.025}    \\
		$zT$  &   MAE &   0.098$\pm$0.009 &   0.099$\pm$0.009    &    0.094$\pm$0.007  &  \textbf{0.088$\pm$0.010}    \\    
  		&   R$^2$  &    0.678$\pm$0.068 &   0.668$\pm$0.055    &    0.708$\pm$0.034    &    \textbf{0.741$\pm$0.065}    \\
		\bottomrule
	\end{tabular}
	\label{tab:table_mmoe_representations}
\end{table}


\begin{figure}[!t]
	\centering
        \includegraphics[width=\columnwidth]{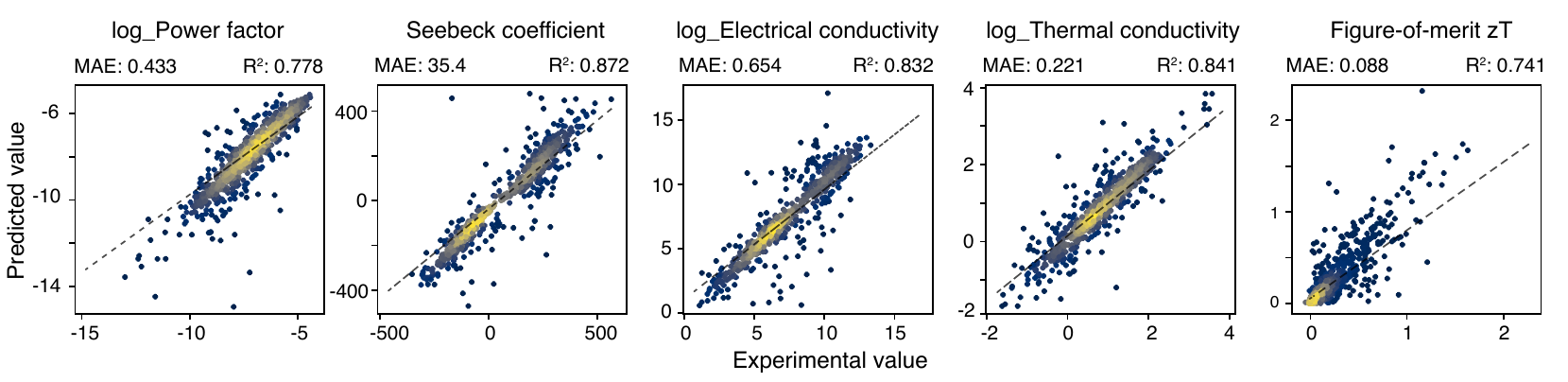}
	\caption{Multi-task prediction results of five TE properties from the best performing MMoE model with 5-fold cross validation on the experimental dataset. Composition embeddings, structure embeddings, and context features are concatenated as the model input. Color indicates density. }
	\label{fig:fig5_mmoe_predictions}
\end{figure}


The accuracy of the property predictions is rooted in the quality of the data representations. 
In addition to the embeddings derived from language models, we added further information based on context features as model input. 
Materials science optimization techniques including doping/defect-engineering 
(\cite{toriyama2021defect}), alloying and phase-boundary mapping (\cite{ohno2018phase, ortiz2019carrier}) are widely utilized and critical to enhance the performance of TE materials. 
The composition of a material after optimization (e.g. doping) is different  from the original composition of the host material via the introduction of dopants and other defects.  
A small degree of doping can substantially affect TE performance. 
For example, in the experimental database the reported $zT$ of PbTe can vary from as low as 0.10 to as high as 1.56 depending on doping/alloying strategy according to the UCSB dataset(\cite{gaultois2013data}). 
Moreover, material properties resulting from different synthesis conditions (especially  temperature) can vary substantially.

For these reasons, we devised different material representations that can include up to three components: (i) host material structural features, (ii) composition features accounting for doping, phase boundary mapping, and alloying (on normalized cheical formulae), and (iii) context features (one-hot encoded temperature).  
For property prediction, structure embeddings carry important information regarding host material structure, while composition embeddings bring in information about off-stoichiometry.  
We summarize how different input material representations affect MMoE performance for TE tasks in Table \ref{tab:table_mmoe_representations}. 
With context features being included, the best performing multi-task TE model was achieved by concatenating doped composition, host structure, embeddings (Table \ref{tab:table_mmoe_representations}), whereas host composition embeddings alone gave the worst performance metrics. 
While doping and alloying often present significant challenges for first-principles modeling, the language representation accounts for such material complexity naturally, through the contextual knowledge contained in the embedding.
In general, models trained with both structure and composition representations perform consistently better than those with only composition embeddings. 
Therefore, modeling TE properties requires accurate representations of both structural and doped compositions, which can be effectively extracted through BERT-based language models. 
The multi-task learning results from our best-performing material representation and MMoE is shown in Figure \ref{fig:fig5_mmoe_predictions}.
In all five prediction tasks, MMoE accurately predicts the TE properties for the input material under each one-hot encoded temperature category with $R^2 > 0.7$. 
Despite being trained directly on general representations of crystals, this model achieves comparable accuracy to recent domain-specific models in the TE field (\cite{na2021predicting, na2022public}).






\newpage
\subsection{Search ranking of TE materials with similar potential}\label{results:ranking}


\begin{figure}[!b]
	\centering
        \includegraphics[width=\columnwidth]{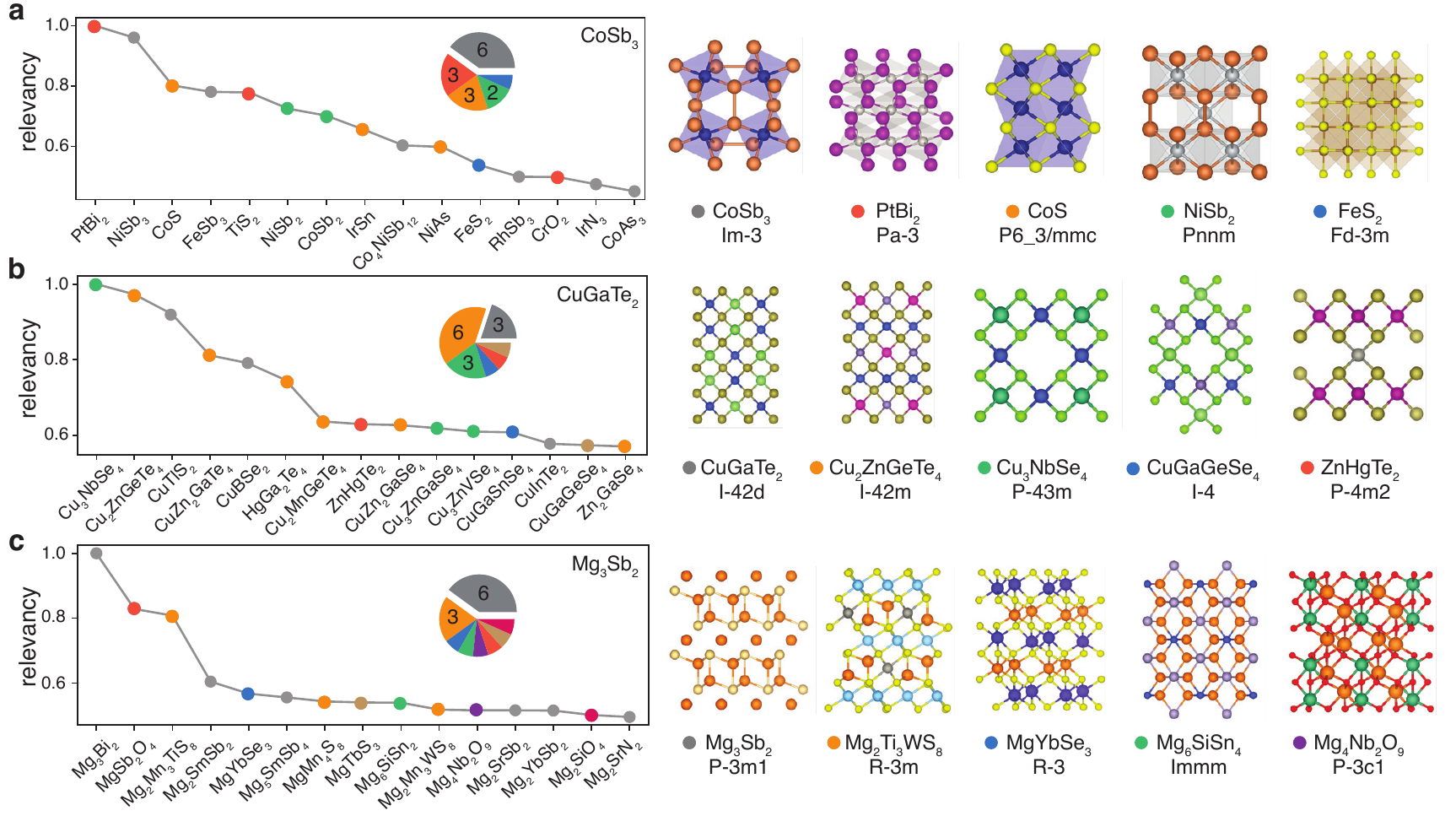}
	\caption{Ranking results of top 15 materials that exhibit most similar TE potential to \textbf{a}, CoSb$_3$, \textbf{b}, CuGaTe$_2$, and \textbf{c}, Mg$_3$Sb$_2$. The color of each data point denotes the structure prototype as shown on the right panel for each query material. The recommended structure prototypes share similar structural features with the query material.}
	\label{fig:fig7_ranking}
\end{figure}

To interpret and evaluate the ranking performance, we demonstrated the ranking outcomes from our recommendation framework on seven representative TE materials. Candidates were ranked by their \textit{relevancy score} (Section \ref{method:evaluation}), which is defined as the reciprocal of the summed absolute percent difference of five properties from the query material. 
Figure \ref{fig:fig7_ranking} shows the ranking results for CoSb$_3$, CuGaTe$_2$, and Mg$_3$Sb$_2$, representing skutterudite, diamond-like semiconductors (DLS), and Zintl phases.
For each query material, the top 15 ranked materials that exhibit the most similar TE potential are shown. A full list of the search ranking results for the other materials (PbTe, BiCuSeO, Cu$_2$Se, Bi$_2$Te$_3$) can be found in Supplementary Figure S6.

In Figure \ref{fig:fig7_ranking}, each candidate is colored by its structure prototype to visualize the structural diversity.
The distribution of prototype structures is shown by the pie chart.
For skutterudite CoSb$_3$, the top 15 recommendations 
consist of 5 prototype structures, and 9 out of the 15 top 
are different from the space group of CoSb$_3$ (I$\bar{m}$3, No. 204).
As expected, several AX$_3$ skutterudites (grey in Figure \ref{fig:fig7_ranking}a) appear in the list, sharing the same prototype structure with CoSb$_3$.
Several novel structure prototypes also appear, including pyrite (P$\bar{a}$3, No. 205, red in Figure \ref{fig:fig7_ranking}a) and marcasite structures (\textit{Pnnm}, No. 58, green in Figure \ref{fig:fig7_ranking}a), as close relevant TE materials to CoSb$_3$. 
Moreover, two more prototype structures --  covellite (P6$_3$/mmc, No. 194, orange) and carrollite-like AX$_2$ structure (F$\bar{d}$3m, No. 227, blue) are also recommended, both of which have received limited attention historically but may warrant further investigation (\cite{mukherjee2019effect}).
Note that all above structure prototypes have corner-sharing octahedral motifs, a local structural feature shared with query material CoSb$_3$ that may correlate to similar TE properties. 
The recommendations based on querying of diamond-like chalcopyrite material CuGaTe$_2$ (I$\bar{4}$2d, No. 122, grey in Figure \ref{fig:fig7_ranking}b) render diversified outcomes with 5 different structure prototypes. 
In addition to four more ABX$_2$ chalcopyrites, the framework selected quaternary stannite (I$\bar{4}$2m, No.  121, orange in Figure \ref{fig:fig7_ranking}b), sulvanite (P$\bar{4}$3m, No. 215, green in Figure \ref{fig:fig7_ranking}b), defect kesterite ($\bar{I}$4, No. 79, blue in Figure \ref{fig:fig7_ranking}b), and chalcopyrite-like (P$\bar{4}$m2, No. 115, red in Figure \ref{fig:fig7_ranking}b) structures.
For Zintl phase Mg$_3$Sb$_2$, the top 15 recommendations comprise 8 unique prototype structures (5 of which are shown in Figure \ref{fig:fig7_ranking}c).
Interestingly, the prototypes do not exhibit the layered structure of query material Mg$_3$Sb$_2$.
Instead, the common local structural feature of octahedral motifs is 
present throughout the recommended prototypes.
Unlike other computational materials discovery strategies which generate candidate materials by applying chemical substitutions to a single prototype structure (\cite{wang2021predicting,qu2020doping}), our framework is able to suggest candidates with diversified structures that are different from, but still related to, the prototype. 
Such capability can offer insights and  understanding of  structural similarity between different prototypes and structure-to-property mappings for ML tasks.

To evaluate the performance of the ranking tasks, we performed first-principles calculations on the TE properties of top recommended candidates (see computational details in \ref{method:AMSET}).
As shown in Figure S10, the calculated properties of the recommended materials resemble those of the query material.
For example, both CuGaTe$_2$ and its top ranked candidates exhibit high $p$-type TE performance that outperforms the $n$-type counterparts.
Upon experimental evaluation of several top ranked candidates, we identified CuZn$_2$GaTe$_4$ as a novel $p$-type TE material with high Seebeck coefficient ($S = 250 \mu$V/K at 575 K) and overall decent TE properties, see Figure S11.  
This immediate positive result arose from self-doping that yielded an optical carrier concentration near 4.5$\times 10^{19}$~cm$^{-3}$ at 473 K. 
Preliminary experimental measurements on other candidates, while not demonstrating good performance immediately, revealed individual features that are beneficial to TEs and the potential to achieve good performance upon further optimization. 
The most important features are, e.g., strongly suppressed thermal conductivities at room temperature of HgGa$_2$Te$_4$ (0.36 W/mK) and CuGaGeSe$_4$ (0.62 W/mK). Cu$_2$ZnGeTe$_4$ also belongs to the group of CuGaTe$_2$-like materials and was previously reported to show decent mobility (ca. 30 cm$^2$/Vs at 550 K, \cite{ortiz2018ultralow}). Lastly, TiS$_2$ suggested as relevant to CoSb$_3$, has previously been reported with large Seebeck and $zT$ = 0.3 at 700 K (\cite{bourges2016thermoelectric}).





\begin{figure}[!b]
	\centering
        \includegraphics[width=\columnwidth]{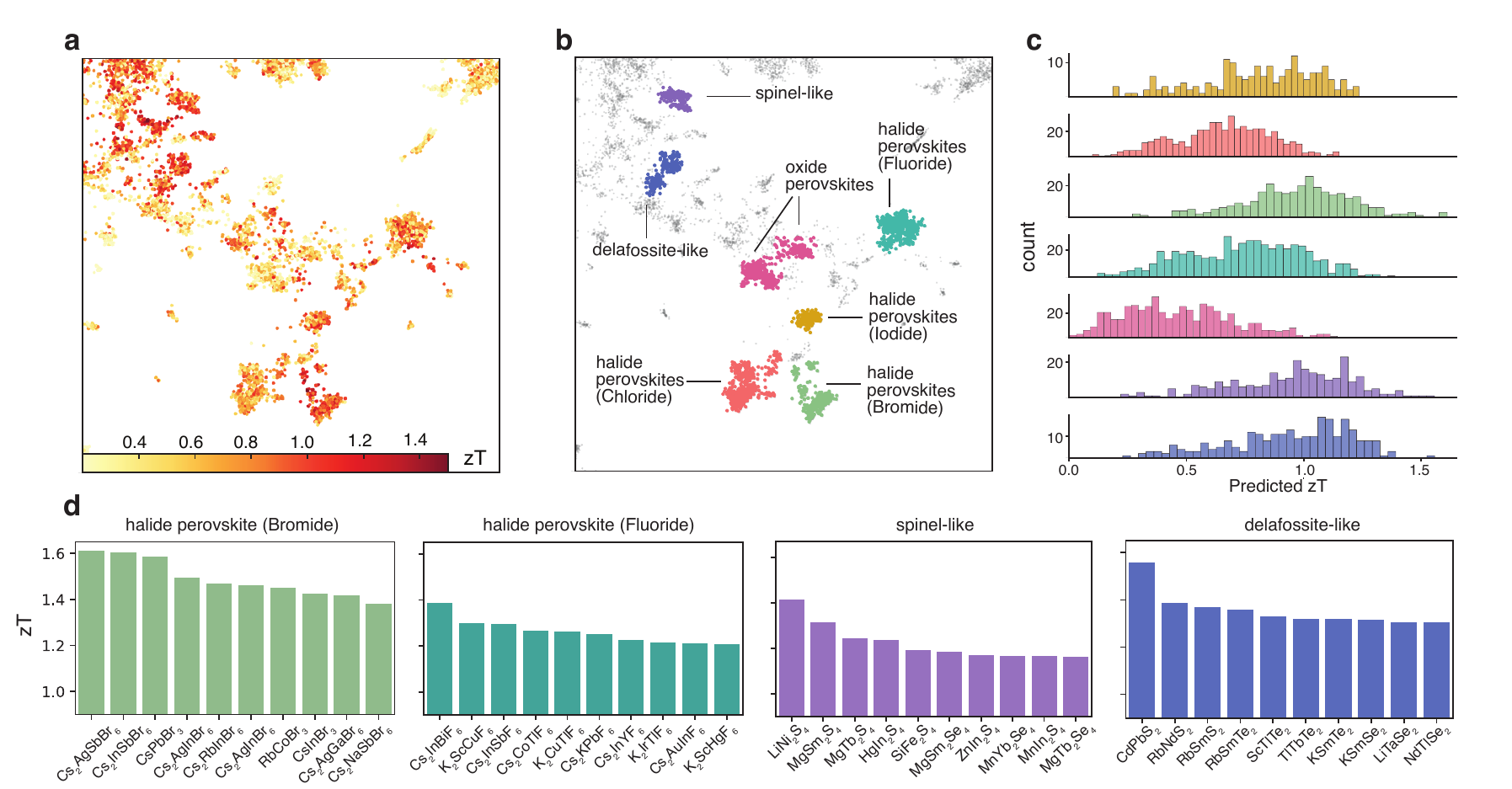}
	\caption{\textbf{a}, Zoomed-in region of UMAP projected embedding space as shown by the dashed line in Figure \ref{fig:fig2_umap_embedding}, colored by predicted $zT$ from the MMoE model. \textbf{b}, Groups of materials that exhibit high predicted $zT$ are colored in the UMAP. \textbf{c}, Distributions of the predicted $zT$ from each material group in \textbf{b}. \textbf{d}, Top 10 candidates ranked by their predicted $zT$ from each material group.} 
	\label{fig:fig8_exploration}
\end{figure}


\subsection{Exploration of under-studied materials in the representation space.}\label{results:exploration}

Through material language representations, we noticed that the distributions of materials for both known and predicted high $zT$ 
appear within the same ``band''-like region 
in the UMAP (Figure \ref{fig:fig2_umap_embedding}d).
Despite their high predicted $zT$, materials at the bottom right corner of the ``band''  (shown by the dashed grey box in Figure \ref{fig:fig2_umap_embedding}) are under-explored with no records from the experimental datasets (Figure \ref{fig:fig8_exploration}a).
In that region, we have identified high-$zT$ clusters composed of halide perovskites (fluoride, chloride, bromide, and iodide), oxide perovskites, spinel-like, and delafossite-like structures, as labelled in Figure \ref{fig:fig8_exploration}b.
The distributions of predicted $zT$ for all materials in each cluster are visualized in Figure \ref{fig:fig8_exploration}c. 
Among halide perovskites, bromides have the highest predicted $zT$ with a mean above 1.0, while fluorides, chlorides, and iodides are close to each other in the predicted $zT$ distributions.
The top 10 highest $zT$ candidates from bromide and fluoride perovskite clusters (Figure \ref{fig:fig8_exploration}d) are mostly Cs- and K-containing double pervoskites A$_2$BB$^{'}$X$_6$, with a few single perovskites ABX$_3$. 
The high TE performance of halide perovskites can likely be attributed to low thermal conductivity. 
\cite{lee2017ultralow} revealed that inorganic halide perovskites exhibit ultra-low thermal conductivity due to a unique cluster rattling mechanism, resulting in thermal conductivities comparable to the amorphous limit. 
One of the top predicted candidates CsPbBr$_3$, in particular, has attracted wide attention in the TE field (\cite{yan2020symmetry}).
Recent first-principles calculations (\cite{mahmood2022study, saeed2022first}) support our findings of high TE performance for several double perovskites. 
It is worth noting that all of the recommended candidates are lead-free and have high-temperature stability (\cite{gao2021screening}), good oxidation resistance, and lower processing costs.
With recent experimental advances in improving the stability of perovskites (\cite{niu2015review, tiep2016recent}), halide perovskites may become more appealing new TE candidates.
Oxide perovksites, interestingly, show inferior TE performance to halide counterparts.
This observation aligns with chemical intuition, since oxides are in general more ionic and insulating materials, rendering them hard to dope to optimize the TE power factor. 

Delafossite-like and spinel-like structures are also under-explored structural spaces with potential to host TE candidates. 
Unlike perovskites which form isolated clusters in the representation space, these two material groups neighbor the more well-explored chalcopyrites and AB$_2$X$_2$ Zintl phases.
Delafossite-like structures, also known as caswellsivlerites, refer to ABX$_2$ (X = O, S, Se, Te) materials crystallizing in the trigonal structure, with CuFeO$_2$ (R$\bar{3}$m, No.166) being the prototype.
So far, extensive research efforts have focused on delafossite-type oxides as TE materials (\cite{hayashi2008effect, van2020effects}), while a recent high-throughput computational study (\cite{shi2017high}) revealed that sulfide, selenide and telluride delafossite-like structures are also thermodynamically stable. 
For the top 10 predicted $zT$ candidates from the delafossite group (Figure \ref{fig:fig8_exploration}d), all candidates are sulfides and selenides that we recommend for further investigation.
Similarly, less attention has been focused on sulfide, selenide, and telluride spinels compared to oxide counterparts, while all top 10 spinels from the recommended list (Figure \ref{fig:fig8_exploration}d) are sulfides and selenides. 
Recent theoretical works have suggested several high-performance spinel sulfides Tm$_2$MgS$_4$ ($zT\sim$0.8, \cite{nazar2022first}), Y$_2$CdS$_4$ ($zT\sim$0.8, \cite{yakhou2019theoretical}), MgIn$_2$Se$_4$ ($zT\sim$0.7, \cite{mahmood2019opto}), suggesting that the discovery framework is able to select  good candidates from a large and diverse search space. 
As evaluation of the recommended under-explored material groups, we performed first-principles calculations on the top candidates from each group in Figure \ref{fig:fig8_exploration}d, which are summarized in Figure S10.
These calculations corroborated the promising TE potential of  several candidates, e.g. delafossite-like CdPbS$_2$ ($n$-type, $zT_\mathrm{max}$=1.7 at 800~K), halide perovskite Cs$_2$InSbF$_6$ ($n$-type, $zT_\mathrm{max}$=1.0 at 800~K), etc.




\section{Discussion}
While representation learning has facilitated extraction of more meaningful features from large unlabeled data, methods for learning material representations have also gained substantial momentum (\cite{xu2021self, gupta2021cross, na2022contrastive}). 
On the other hand, language-based models have achieved remarkable outcomes in prediction and generation tasks across an extensive array of domain areas. 
In this work, we demonstrated the use of language representations in the inorganic crystalline materials domain.
Specifically, we introduced a language-based framework to extract composition and structure embeddings as material representations via pretrained language models. The discovery framework is designed to be task-agnostic. We anticipate that it can be expanded upon and utilized to search and explore vast chemical and structural spaces, towards functional materials design and discovery. 

Representing materials in the format of natural language enables effective utilization of material science knowledge learnt from ever-growing unstructured scientific texts. 
Indeed, the extracted embeddings form a chemically meaningful representation space without task-specific supervision. We find that knowledge can be extracted from representations by unsupervised recall on embedding vectors and supervised neural networks, together enabling the funnel-based approach.
In particular, the recall step allows reliable recommendation by constraining the ranking on candidates that are similar to the query material in the representation space.
A benefit of such pre-screening is the avoidance of common pitfalls where materials exhibit similar properties that arise for inherently different reasons, i.e. far from each other in the representation space. 
For the use case of thermoelectrics, for example, high $zT$ can arise from either high power factor or low thermal conductivity. 
Another strength of language representations is that they can effectively handle off-stoichiometric material compositions to account for alloying and doping, which typically require complicated computational techniques (e.g. disorder modelling) for accurate predictions in first-principles simulations.

Exploitation and exploration trade-off has been a common phenomenon in recommender systems (\cite{gao2021advances, vanchinathan2014explore}).
For our recommendation framework, while exploitation refers to seeking maximum reward, exploration may be thought of as consideration of new structural prototypes present in the top-ranked candidates that share structural features with, but are distinct from, the query material. 
A reasonable balance between exploitation and exploration, which can be tuned by the number of candidates recalled from the candidate generation step, will diversify the recommendation while still proposing structurally-related materials.
For example, the top-15 ranked materials for both CoSb$_3$ and CuGaTe$_2$ contain 5 different prototype structures when 100 recalled materials from the candidate generation step are considered for ranking, 
while the number of prototypes increases to 14 and 9 respectively if the number of recalled materials considered is increased to 1000 (Supplementary Figure S7).

As future directions for language representation for crystals, we suggest to enrich the material representations by diversifying both text-based input and structures.
The automatically generated text descriptions from Robocrystallographer 
are monotonic with little variation between descriptive words/phrasing (\cite{sayeed2023structure}). 
These descriptions can possibly be  diversified via paraphrasing or developing structure to sentence machine translation models to describe crystal structures in text.
On the other hand, the structural complexity in the representation space can be diversified via generative models, e.g., diffusion models (\cite{xie2021crystal, lyngby2022data}), to design new prototype structures beyond simple lattice decoration of known crystals.


\section{Methods}
\label{sec:headings}

\subsection{Data preparation}\label{method:dataset}

The training dataset was collected from the Materials Project (\cite{jain2013commentary}) to include 116,216 materials are likely to be thermodynamically stable.
Using decomposition enthalpy < 0.5 eV as a query criteria, we utilized Materials Project API (\cite{ong2015materials}) and Pymatgen (\cite{ong2013python}) library to collect materials for use in this study.

In this work we considered five different datasets, all of which includes properties relevant to thermoelectric materials; UCSB dataset (\cite{gaultois2013data}) -- an experimental dataset from Materials Research Laboratory (MRL) about 1092 materials (500 unique materials) with their thermoelectric properties;
ESTM dataset (\cite{na2022public}) -- an experimental dataset containing 5205 materials (880 unique materials) with their thermoelectric properties; 
ChemExtracter dataset (\cite{sierepeklis2022thermoelectric}) -- a mixture of experimental and theory dataset by auto-generation from the scientific literature spanning 10,641 unique chemical names;
TEDesignLab dataset (\cite{gorai2016te}) -- a theory dataset containing lattice thermal conductivity for 3278 materials;
Citrine dataset (\cite{ward2018matminer}) -- an experimental dataset from Matminer (\cite{ward2018matminer}) containing thermal conductivity records for 871 materials.
In all five datasets, 826 materials that have records for five TE properties are used for evaluation of recall performance in Section \ref{method:evaluation}.
We calculated the numeric mean for materials with repeated entry for certain properties and properties at different temperatures.
For MMoE model training and testing, UCSB and ESTM dataset are utilized as ground-truth labels .
During training, the TE properties are matched to corresponding temperature range via one-hot encoding.

\subsection{Embedding models}\label{method:embedding_models}
Three model-based and one model-free embedding methods were used in this work. 
For the model-based approach, we obtained pretrained weights for Mat2Vec (\cite{tshitoyan2019unsupervised}), MatsciBERT (\cite{gupta2022matscibert}), and MatBERT (\cite{trewartha2022quantifying}). 
For Mat2Vec, it was trained similarly as Word2vec training through skip-gram with negative sampling. 
Each word is embedded into a 200-dimensional vector. For the BERT-based models, MatsciBERT was pretrained on whole sections of more than 1 million material science articles, whereas MatBERT was trained by sampling 50 million paragraphs from 2 million articles. 
Both models were trained with masked language modeling (15\% dynamic whole word masking) and next-sentence prediction as the unsupervised training objectives. 
Both models are uncased, and have maximum 512 input token size with 768 hidden dimensions. 
The vocabulary size for the tokenizer is 30,522. 
For the fingerprint generation, it was generated using CrystalNN (\cite{zimmermann2020local}) algorithm as implemented in Matminer (\cite{ward2018matminer}) package.
The fingerprint contains statistical information about local motifs with a size dimension of 122.

\subsection{Material language representations}\label{method:pretrained_languagemodels}
We acquired compositional and structural level representations for 116K materials in total. 
To acquire structural level representations for each individual material, we applied robocrystallographer (\cite{ganose2019robocrystallographer}), an open-source toolkit that converts the material structure into a human-readable text passage describing local, semi-local and global structural features of the given material.
We used robocrystallographer descriptions from \cite{sayeed2023structure}.
Similar to material descriptions found in literature, such material passage encodes naturally interpretable structural information. 
The whole passage is processed by tokenizers and fed into the pretrained BERT models (MatsciBERT and MatBERT) for output embeddings from hidden layers. 
The output embeddings are $L$ by 768 dimensional matrix, where $L\in$(0,512] is the total number of tokens within the passage. 
We partitioned passages with more than 512 tokens to fit the maximum input token size. 
The final embeddings for each material are constructed by averaging output embeddings across all tokens, resulting in a fixed length of vector representations with 768 dimensions.

For the compositional level representations, Mat2Vec embeddings are directly obtained as the 200-dimensional word embedding vectors of the material formulas. 
With BERT models, we performed same tokenization and embedding procedures on material formulas only. 
This results in the same number of 768-dimensional embedding vectors but only contains information related to the material composition. 
For composition embeddings of the doped material formulas (UCSB dataset), we normalized the compositions to the element with the most number of atoms in the unit cell. 
The output embeddings are obtained on the normalized formulas.

\subsection{MMoE and TE property prediction}\label{method:mmoe}
A shared-bottom multi-task network was first introduced by \cite{caruana1997multitask} and widely applied for multi-task learning. The basic network formulation is: 
\begin{equation}\label{equ:shared-bottom}
    y_{k} = h^{k}(f(x)) 
\end{equation}
where $k=1,2,3...K$ for $K$ number of tasks, $f$ is the shared-bottom network, $h^{k}$ is the tower network for task $k$, and $y_k$ is the output for task $k$. The key difference in MMoE network is to substitute the shared-bottom $f$ with MoE layer $f^{k}(x)$ for a specific task $k$, which is defined as:
\begin{equation}\label{equ:moe}
    f^{k}(x)= \Sigma^n_{i=1}g^{k}(x)_{i}f_{i}(x)
\end{equation}
\begin{equation}\label{equ:gate}
    g^{k}(x)= \mathrm{softmax}(W_{gk}x)
\end{equation}
where $i$ = 1, 2, 3... $n$ for $n$ number of experts, $g^{k}(x)$ is the gating network for each task $k$, and $W_{gk}$ is the trainable matrix. In our implementation, all expert network is a three-layered MLP with 128, 64, and 32 dimensions. The gating network is a two-layered MLP with 32 and 16 dimensions. In all of our experiments, networks are trained for 500 epochs with learning rate = $10^{-3}$, weight decay = $10^{-5}$, and batch size=64. We used k-fold cross-validation method to train and evaluate the model performance. For all datasets, we employed 5-fold cross validation by splitting the dataset into 5 nonoverlapping portions. The number of experts is set to 8 for both AFLOW benchmark dataset and TE dataset.

Recent works (\cite{na2022public, na2021predicting}) reported that doping and alloying information, as well as context features greatly enhance the model performance for TE predictions. As for context features for MMoE, we first sorted the continuous temperatures into four ranges (0, 300], (300, 600], (600, 900], (900, $+\infty$], which were one-hot encoded into sparse feature vectors and passed to embedding layers of the MMoE model. Since our structure embeddings are restricted to the host materials, dopant or alloying information will be derived from composition embeddings to delineate the compositional effect. 
To match doped materials to their hosts, we encoded the normalized doped formulas into composition vectors (sparse vector with number of corresponding elements at each site), followed by mapping to existing host composition vectors via cosine similarity. 
Host materials with the highest cosine similarities were selected.


\subsection{Ranking score and exploratory analysis}\label{method:evaluation}
Once candidates are recalled for the query, their predicted properties are used to compute total absolute percent difference (TAPD) defined as: 
\begin{equation}\label{equ:TAPD}
    \mathrm{TAPD} = \Sigma^K_{k=1} \left( \frac{|y_k^{c}-y_k^{q}|}{y_k^{q}} \right)
\end{equation}
where $K$ is the total number of material properties, $y^c$ and $y^q$ are the candidate and query properties respectively. This measures the composite deviation of candidate properties from the query properties. All properties need to be close to those of the query to have a low TAPD. We define \textit{relevancy score} as the reciprocal of TAPD: 
\begin{equation}\label{equ:relevancy}
    \textrm{relevancy} =\frac{1}{\mathrm{TAPD}}
\end{equation}
In our experiments, 100 candidates were recalled per query material. We ranked the candidates based on their \textit{relevancy score}. The scores presented in the figure were normalized by the maximum score within the recalled list. For the exploratory analysis, clusters were hand-selected based on localization of materials with high predicted $zT$. Within each selected cluster, we extracted and ranked the materials according to their $zT$. All predictions were made at high temperature (900, $+\infty$] as the context features. 


\subsection{Evaluation}

\subsubsection{Unsupervised recall of relevant materials}\label{method:reall_metrics}
Recalling relevant material candidates is an unsupervised process which does not require training labels. First, candidates are searched in the representation space by computing cosine similarities between the embedding vector of the query and the rest of the embedding vectors. The similarity-sorted top candidates are returned as the relevant materials. Metrics including Precision$@k$ and Normalized Discounted Cumulative Gain (nDCG) are used to evaluate the recall performance. Such evaluation metrics are common for recommender system, where the goal is to maximize the number of relevant items in the recalled list, i.e., the top$@$k items with $k$ being the size of the list, as well as the relative order of recalled items. 
Precision$@k$ measures the percentage of the relevant materials in the first $k$ recalled materials:
\begin{equation}\label{equ:precision}
    \textrm{precision}@k = \frac{\textrm{relevant items}@k}{k}
\end{equation}
while nDCG is an evaluation method which compares the ideal ranking of a test set (iDCG), with the ranking assigned by the recommendation algorithm (DCG -- Equ.\ref{equ:DCG}).
\begin{equation}\label{equ:DCG}
    \textrm{DCG} = \Sigma^n_{i=1}\frac{\textrm{relevance}}{\log_2(i+1)}
\end{equation}
\begin{equation}\label{equ:nDCG}
    \textrm{nDCG} = \frac{\textrm{DCG}}{\textrm{iDCG}}
\end{equation}

\subsubsection{First-principles calculations}\label{method:AMSET}
The \textit{ab initio} scattering and transport (AMSET, \cite{ganose2021efficient}) software package was used to estimate scattering rates (or lifetime) and transport properties based on momentum relaxation time approximation (MRTA), which has been shown to give comparable results to state-of-art EPW code (\cite{ponce2016epw}). 
The carrier mobility was simulated by considering three scattering processes, including acoustic phonon scattering (ADP), polar optical phonon scattering (POP), and ionized impurity scattering (IMP). 
Each component of carrier lifetime was evaluated by Fermi’s golden rule, with total characteristic scattering time following Matthiessen’s rule. 
The associated Seebeck coefficient, electrical conductivity, and electronic component of the thermal conductivity were were calculated by solving the Boltzmann transport equation (BoltzTraP) using Onsager transport coefficients. 
All \textit{ab initio} inputs are computed from density functional theory (DFT) using the GGA-PBE (\cite{GGA-PBE}) exchange-correlation functional. 
Lattice thermal conductivity ($\kappa_\mathrm{L}$) was calculated using a semi-empirical model based on a modified Debye-Callaway model (\cite{Miller2017}) which captures anharmonicity. 
Bulk modulus ($B$) was determined by fitting the Birch-Murnaghan equation of state to a set of total energies computed at different volumes that were expanded and contracted around the equilibrium volume. 
Other parameters of the semi-empirical model are directly accessible from the relaxed structures, including density, average atomic mass, volume per atom, average coordination number, and number of atoms in the primitive cell. 
The expression for lattice thermal conductivity is give by 
\begin{equation}
    \kappa_\mathrm{L,ac} = A_2 \frac{\Bar{M} v_\mathrm{s}^y}{T V^{2/3}n^{1/3}} + A_3 \frac{v_\mathrm{s}}{V^{2/3}}(1-\frac{1}{n^{2/3}}) \hspace{0.5em}, 
\end{equation}
\noindent where $A_1$ and $A_2$ are fitted parameters, $\bar{M}$ is the average atomic mass, $v_\mathrm{s}$ is the speed of sound, $T$ is the temperature, $V$ is the volume per atom,
and $n$ is the number of atoms in the primitive cell. 
$v_\mathrm{s}$ is approximated as $v_\mathrm{s}$ $\sim$ $(B/d)^{\frac{1}{2}}$.

\subsubsection{Experiments}\label{method:exp}

CuZn$_2$GaTe$_4$, CuGaGeSe$_4$, and HgGa$_2$Te$_4$ samples were prepared from elements: Cu (99.9\%), Hg (99.999\%), Ga (99.999\%), Zn (99.999\%), Ge (99.999\%), and Te (99.999\%), Se (99.999\%). The stoichiometric weights were first sealed in evacuated silica ampoules and melted at 1000$^o$C for several hours. Next, the ingots were milled in high-energy mechanical mill Spex 8000D for 90 min in an inert environment.
The powders were consolidated in an induction heating hot press at 500$^o$ C, 40 MPa for at least 2 hours. Electrical resistivity and Hall coefficient were studied under vacuum on a home-built apparatus with van-der Pauw geometry (\cite{borup2012measurement}). Seebeck coefficient measurements were carried out using a custom-built device (\cite{iwanaga2011high}) in 300 Torr of nitrogen gas. Diffusivity coefficient ($D$) measurements were performed on Netzsch LFA 467 apparatus. To obtain thermal conductivity ($\kappa$), we used formula $\kappa$ = $D C_p d_\mathrm{exp}$, where $C_p$ is heat capacity and $d_\mathrm{exp}$ is experimental density. Values of $C_p$ were obtained from Dulong-Petit law, while density of the samples was measured with geometric method. For all obtained materials $d_\mathrm{exp}$ was ca. 90\% of the theoretical value or higher.

\section*{Data Availability}
The preprocessed AFLOW and thermoelectric datasets used for training and testing the models, as well as material embeddings obtained in this work, are available at \url{https://doi.org/10.6084/m9.figshare.22718668.v1}. 

\section*{Code Availability}
The code and the model weights are available under the MIT license at: \url{https://github.com/ertekin-research-group/Material_Recommender}

\section*{Acknowledgement}
This work was funded with support from the U.S. National Science Foundation (NSF) via Grant No. 2118201 (HDR Institute for Data-Driven Dynamical Design) and Grant No. 1922758 (DIGI-MAT). This work used the Extreme Science and Engineering Discovery Environment (XSEDE) Bridges-2 at the Pittsburgh Supercomputing Center through allocation TG-MAT220011P.

\bibliographystyle{unsrtnat}
\bibliography{paper}  






\newpage
\section*{Supplementary Information}
\setcounter{figure}{0}
\renewcommand{\thefigure}{S\arabic{figure}}

\textbf{Model performance benchmark on small datasets of materials properties}

\begin{figure}[!h]
\centering
\includegraphics[width=0.6\linewidth]{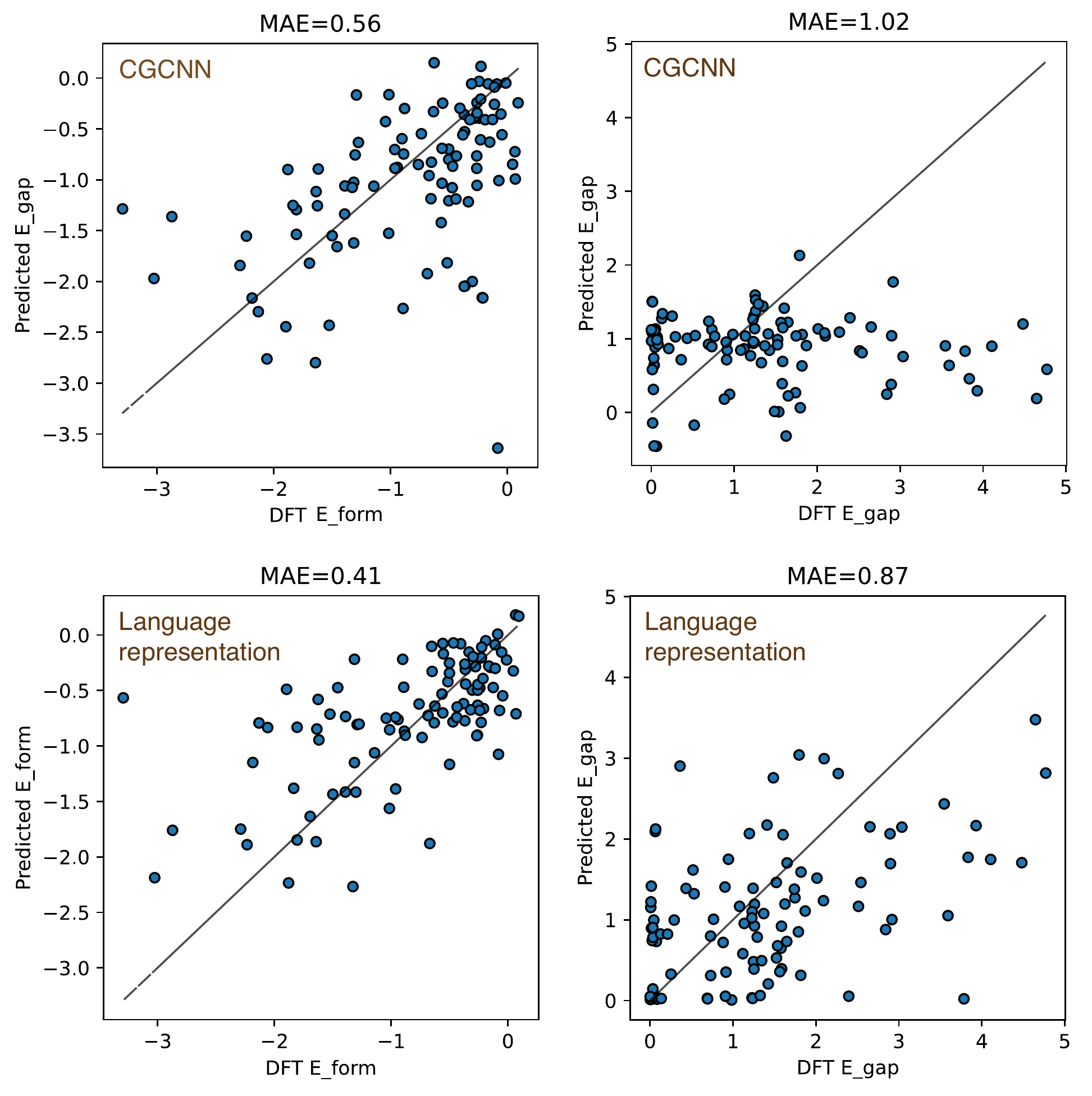}
\caption{Model performance on a small dataset containing 200 training and 100 testing materials for band gap and formation energy predictions. Scatter plots show the comparisons between CGCNN and our models trained with language representations as input.}
\end{figure}

\newpage
\textbf{Single-task predictions on general material properties}

\begin{figure}[!h]
\centering
\includegraphics[width=\linewidth]{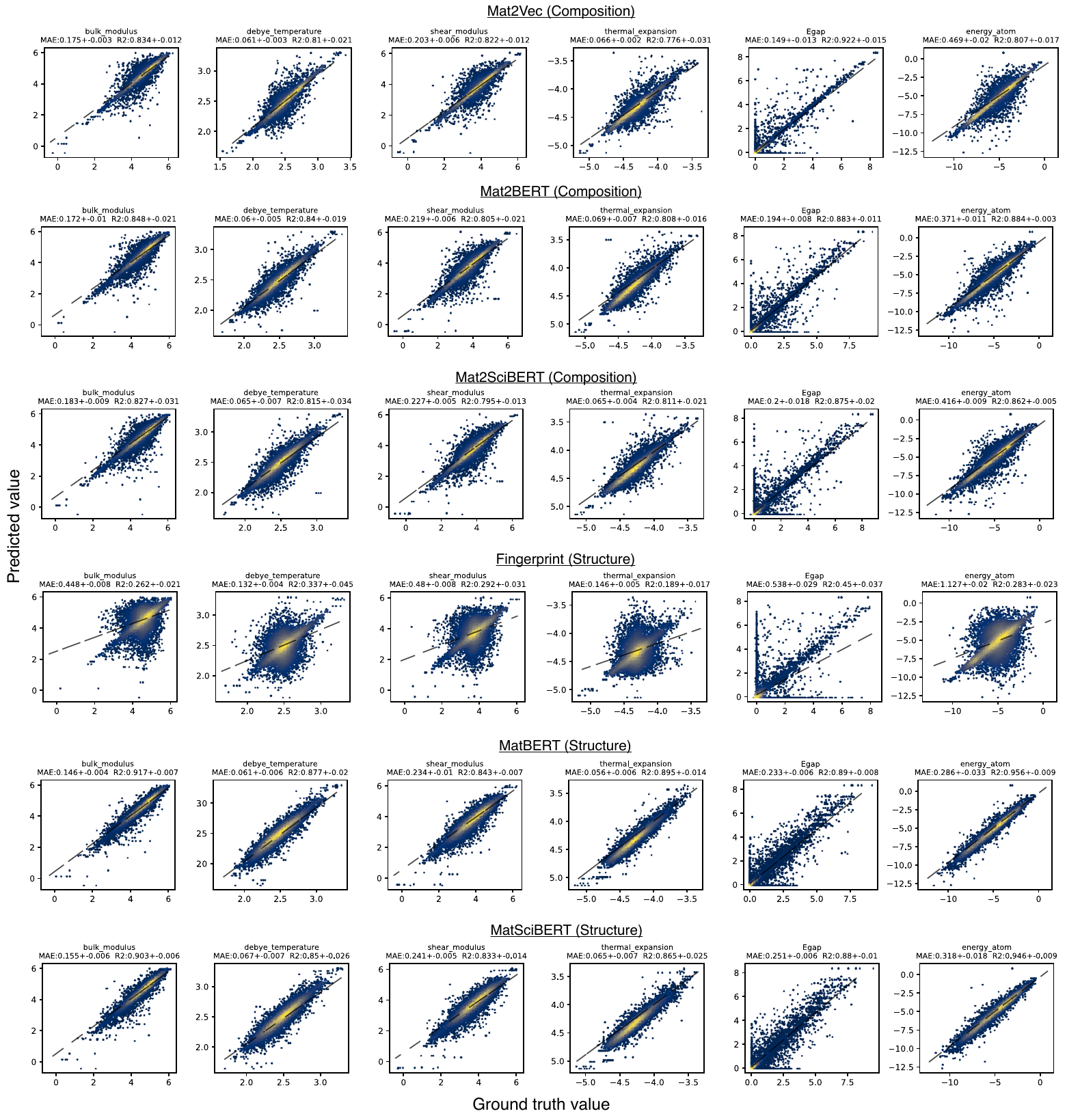}
\caption{
Performance of models trained on individual tasks, evaluated by 5-fold cross validation. Rows correspond to different embedding methods used for model input. Scatter plots show groud truth vs. predicted values on six properties of 5,700 materials from AFLOW dataset.}
\end{figure}

\newpage
\textbf{Multi-task learning on general material properties}

\begin{figure}[!h]
\centering
\includegraphics[width=\linewidth]{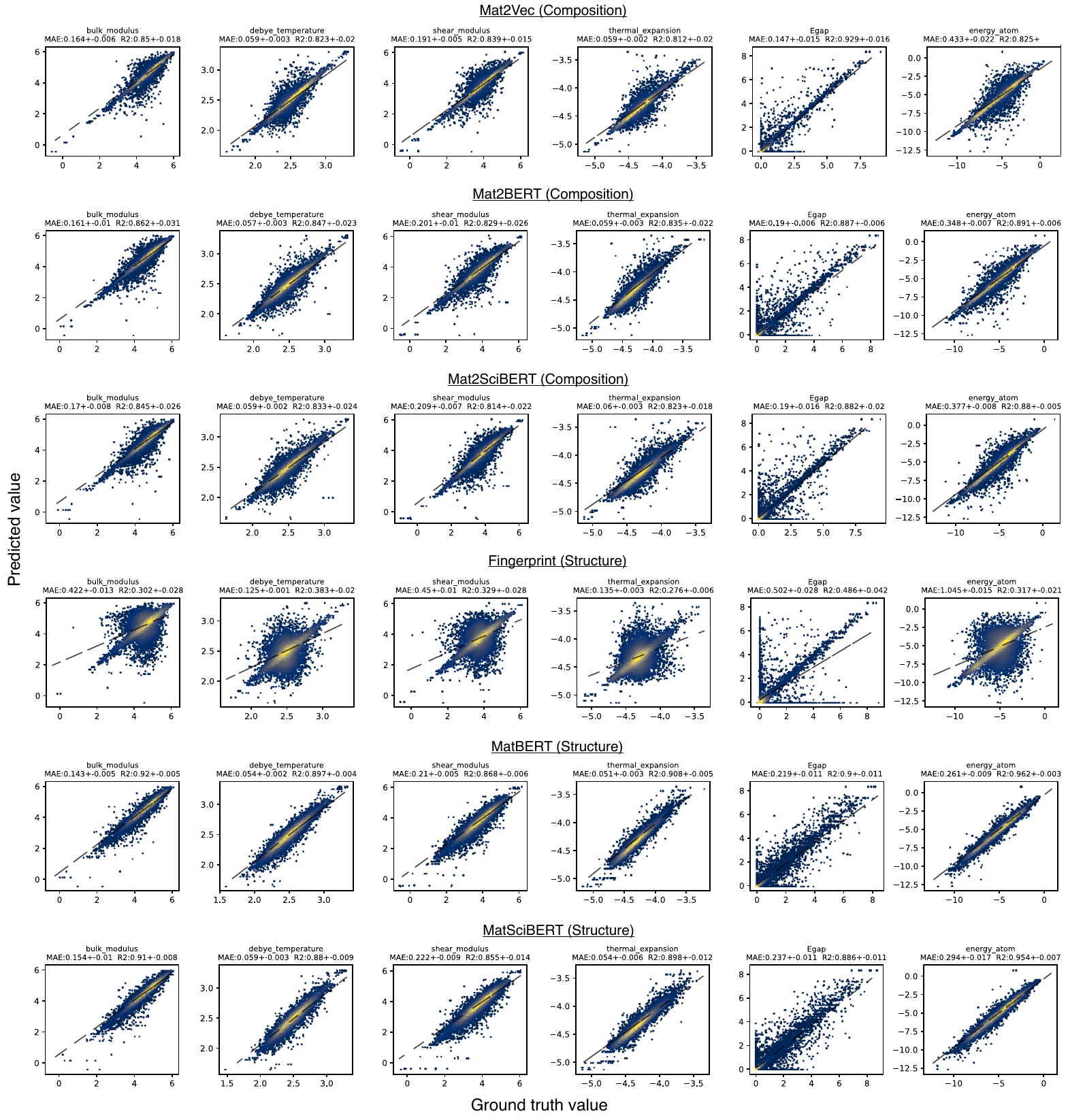}
\caption{Multi-task MMoE model performance evaluated by 5-fold cross validation. Scatter plots show groud truth vs. predicted values on six properties of 5,700 materials from AFLOW dataset. }
\end{figure}

\newpage
\textbf{Correlation between thermoelectric (TE) properties for MMoE prediction tasks}

\begin{figure}[!h]
\centering
\includegraphics[width=\linewidth]{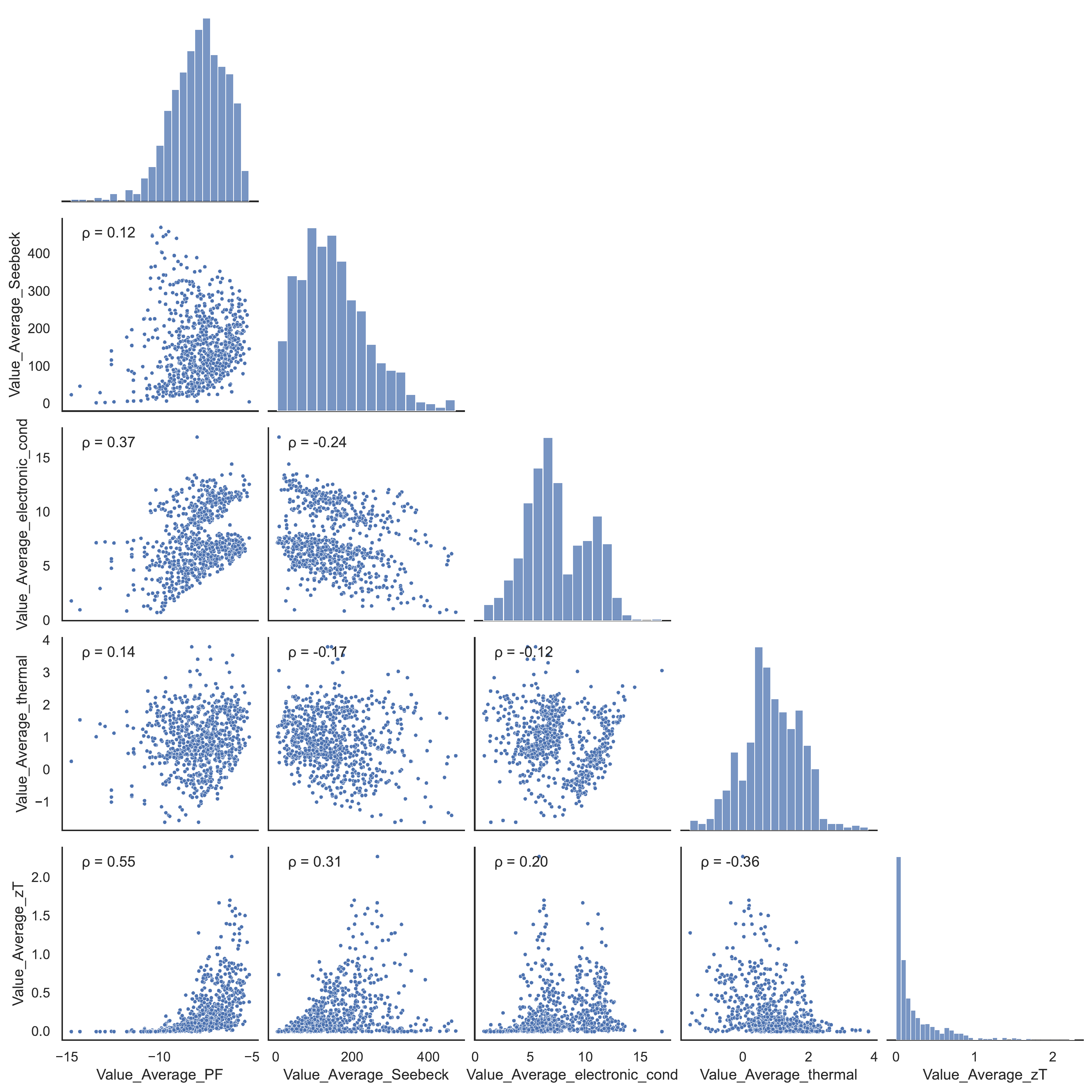}
\caption{ Correlation for five TE properties of the experimental UCSB and ESTM datasets, as indicated by their pairwise Pearson correlation scores ($\rho$). An average of $\sim$0.3 pairwise Pearson correlation presents strong evidence for MMoE to leverage task correlations.}
\end{figure}

\newpage
\textbf{Single task prediction vs. multi-task learning on TE properties}

\begin{figure}[!h]
\centering
\includegraphics[width=\linewidth]{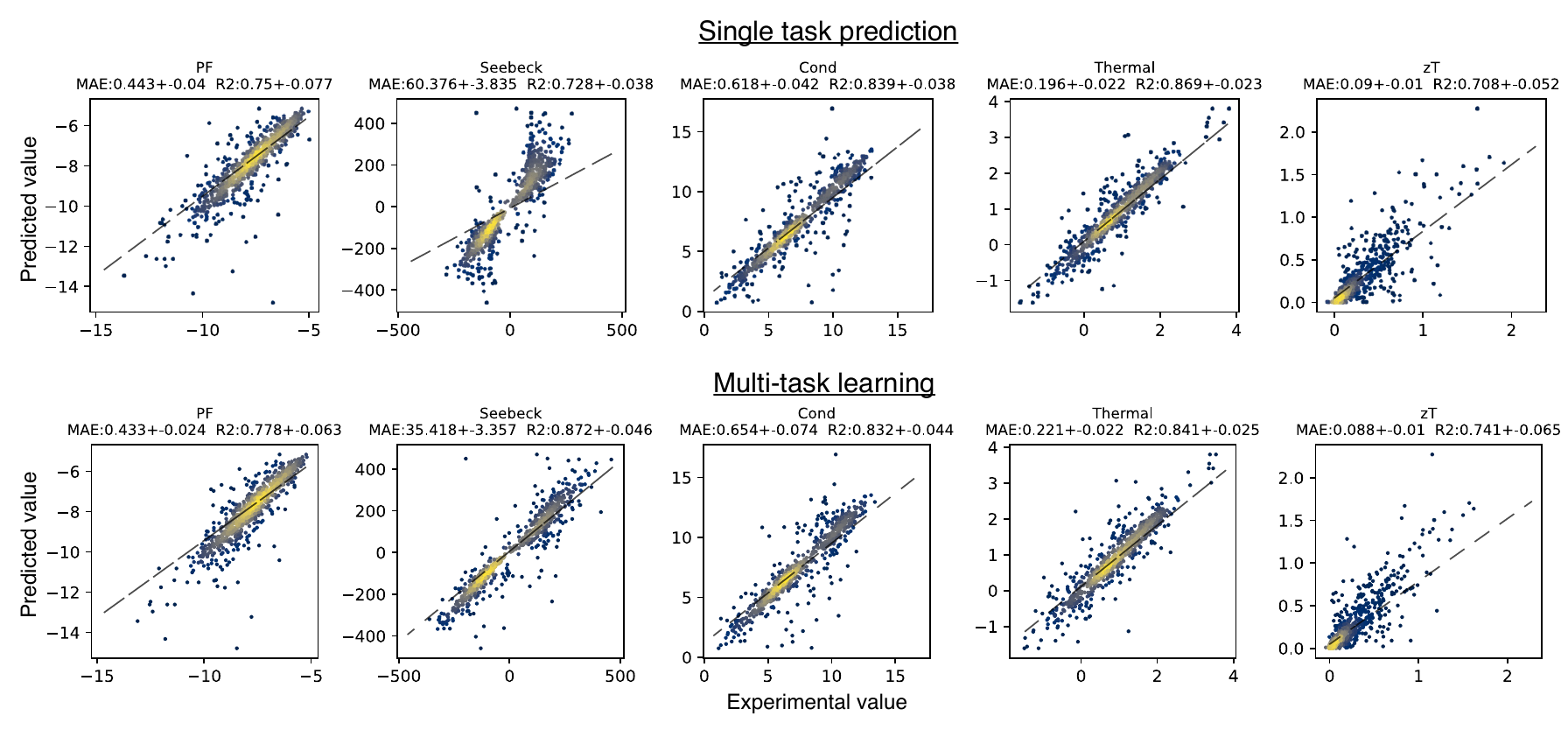}
\caption{ Comparisons of single-task learning (top row) and multi-task learning (bottom row) on TE properties.}
\end{figure}

\newpage
\textbf{Ranking results of top 15 materials (recall top-100 candidates)}

\begin{figure}[!h]
\centering
\includegraphics[width=\linewidth]{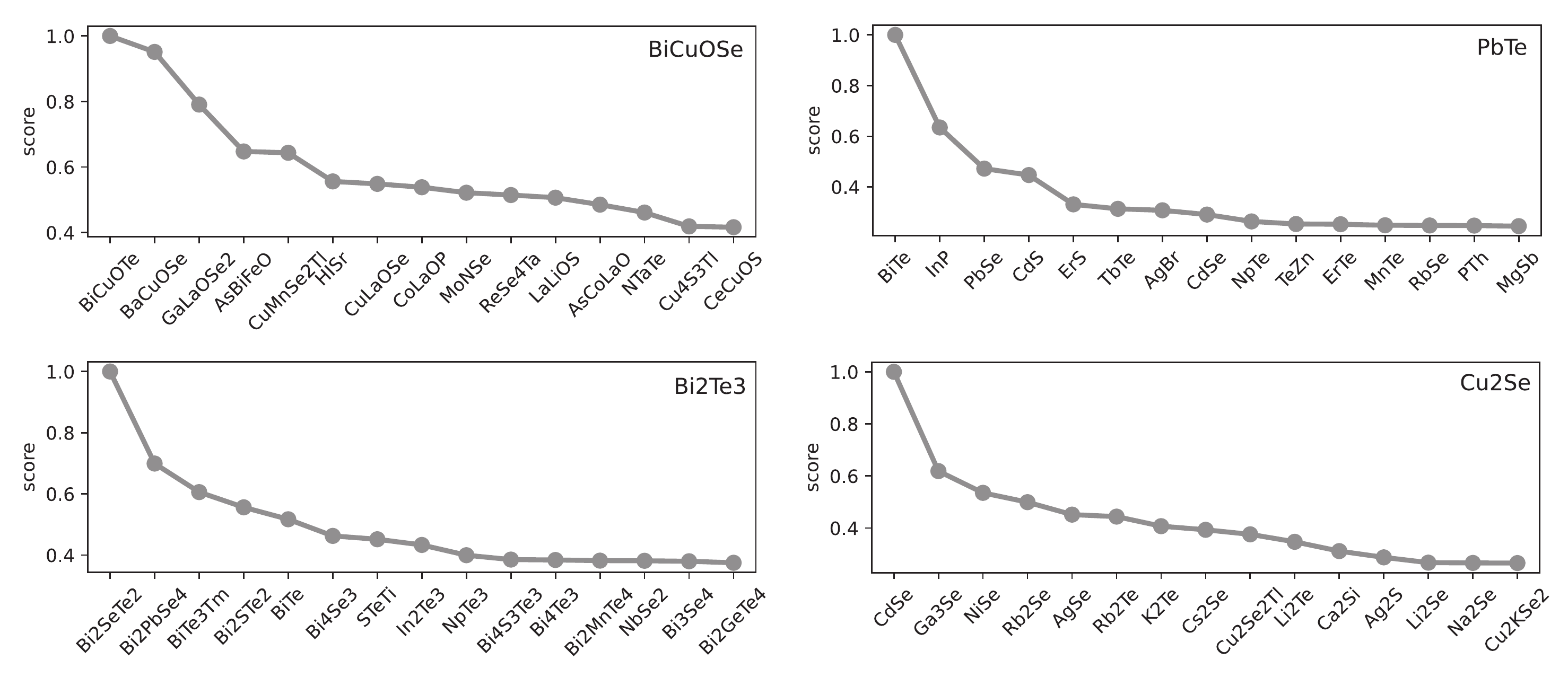}
\caption{
Top-15 candidates from the ranking results of PbTe, Cu$_{2}$Se, BiCuOSe, Mg$_{3}$Sb$_{2}$, and Bi$_{2}$Te$_{3}$.}
\end{figure}

\textbf{Ranking results of top 15 materials (recall top-1000 candidates)}

\begin{figure}[!h]
\centering
\includegraphics[width=\linewidth]{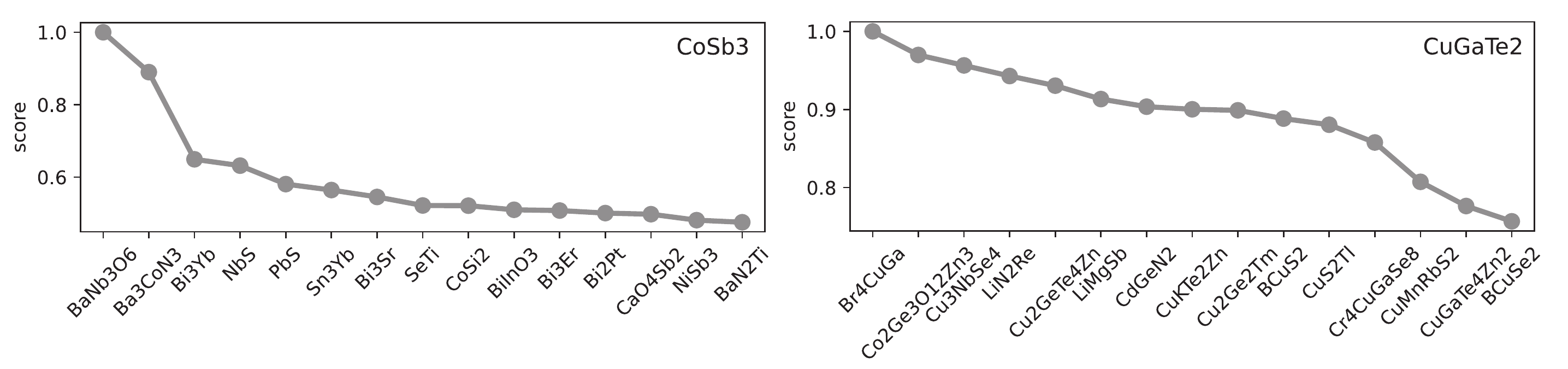}
\caption{
Ranking results of CoSb$_3$ and CuGaTe$_2$ by setting the number of recalled candidates to 1000. More diverse structural prototypes can be seen in the top-15 lists.}
\end{figure}

\newpage
\textbf{Predicted $zT$ distribution of recalled materials for well-known TE materials}

\begin{figure}[!h]
\centering
\includegraphics[width=\linewidth]{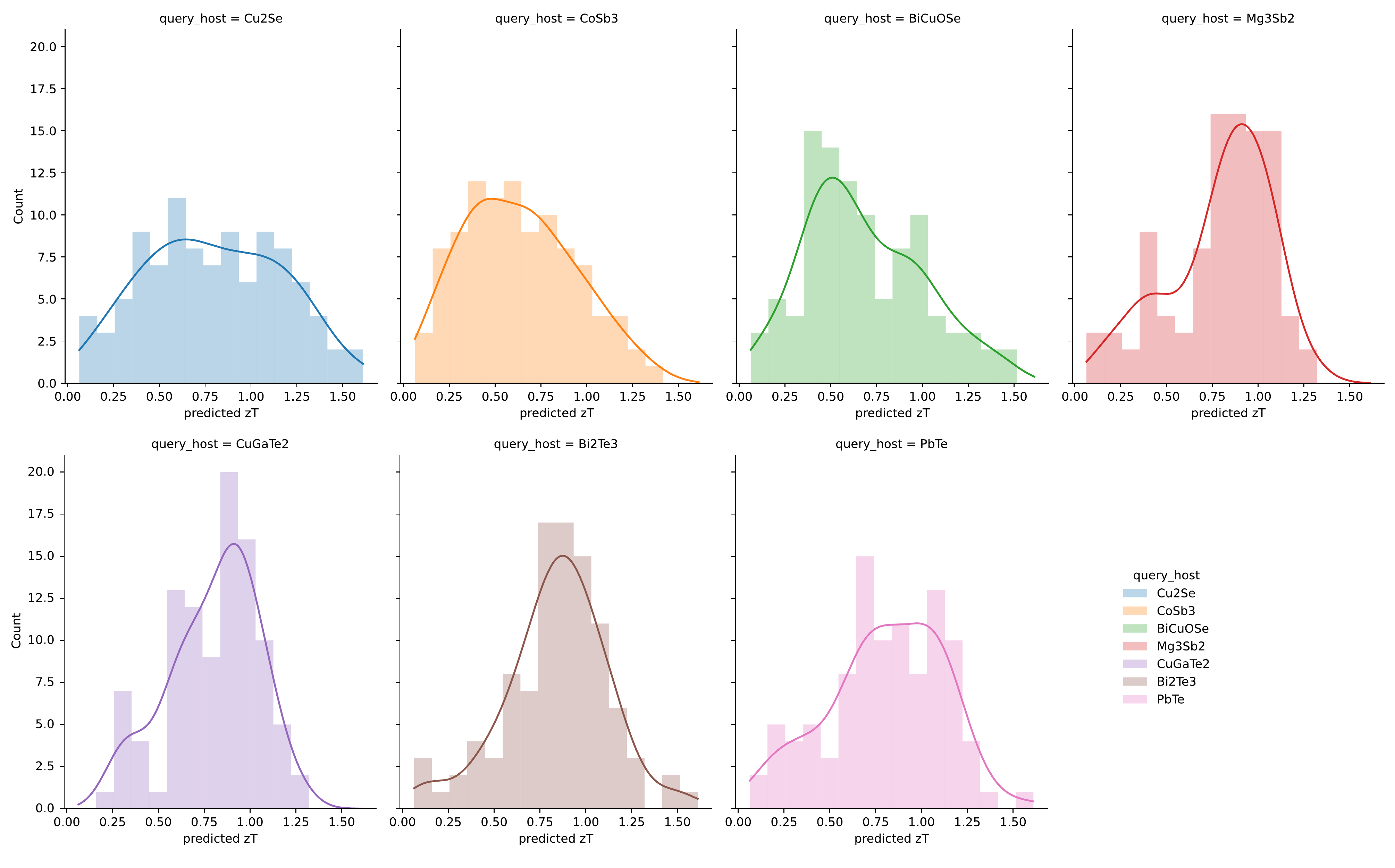}
\caption{Predicted $zT$ distributions of 100 recalled candidates for seven representative TE materials when they are used as query material for the recommendation framework.}
\end{figure}

\textbf{Exploration: Top 10 candidates ranked by their predicted $zT$}

\begin{figure}[!h]
\centering
\includegraphics[width=\linewidth]{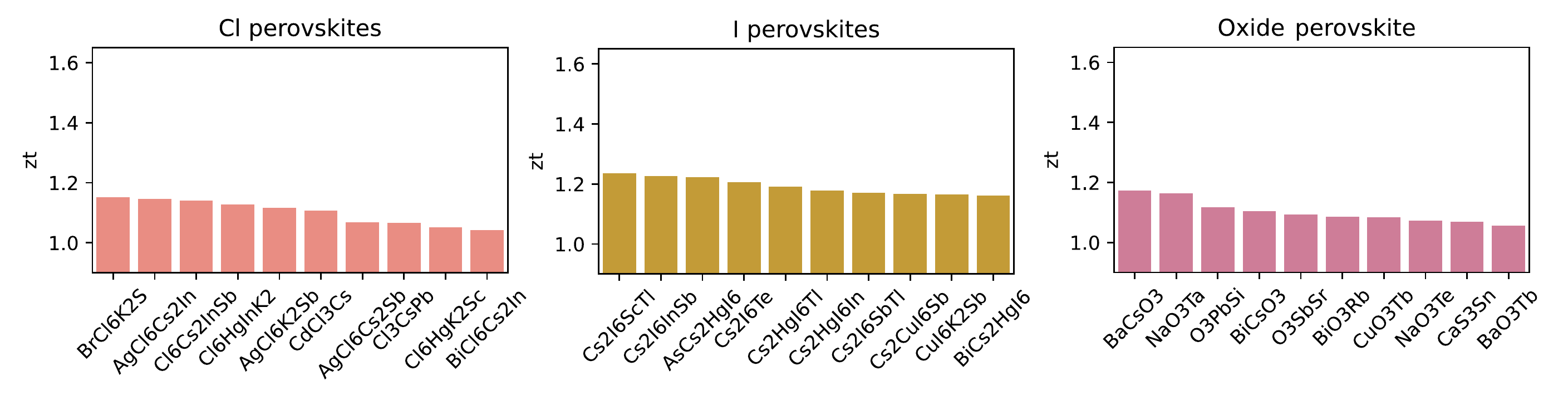}
\caption{
Top-10 ranked candidates with the highest predicted $zT$ in Cl, I, and oxide perovskite groups.}
\end{figure}

\newpage
\textbf{Evaluation: First-principles calculations of TE properties}
\begin{figure}[!h]
\centering
\includegraphics[width=0.95\linewidth]{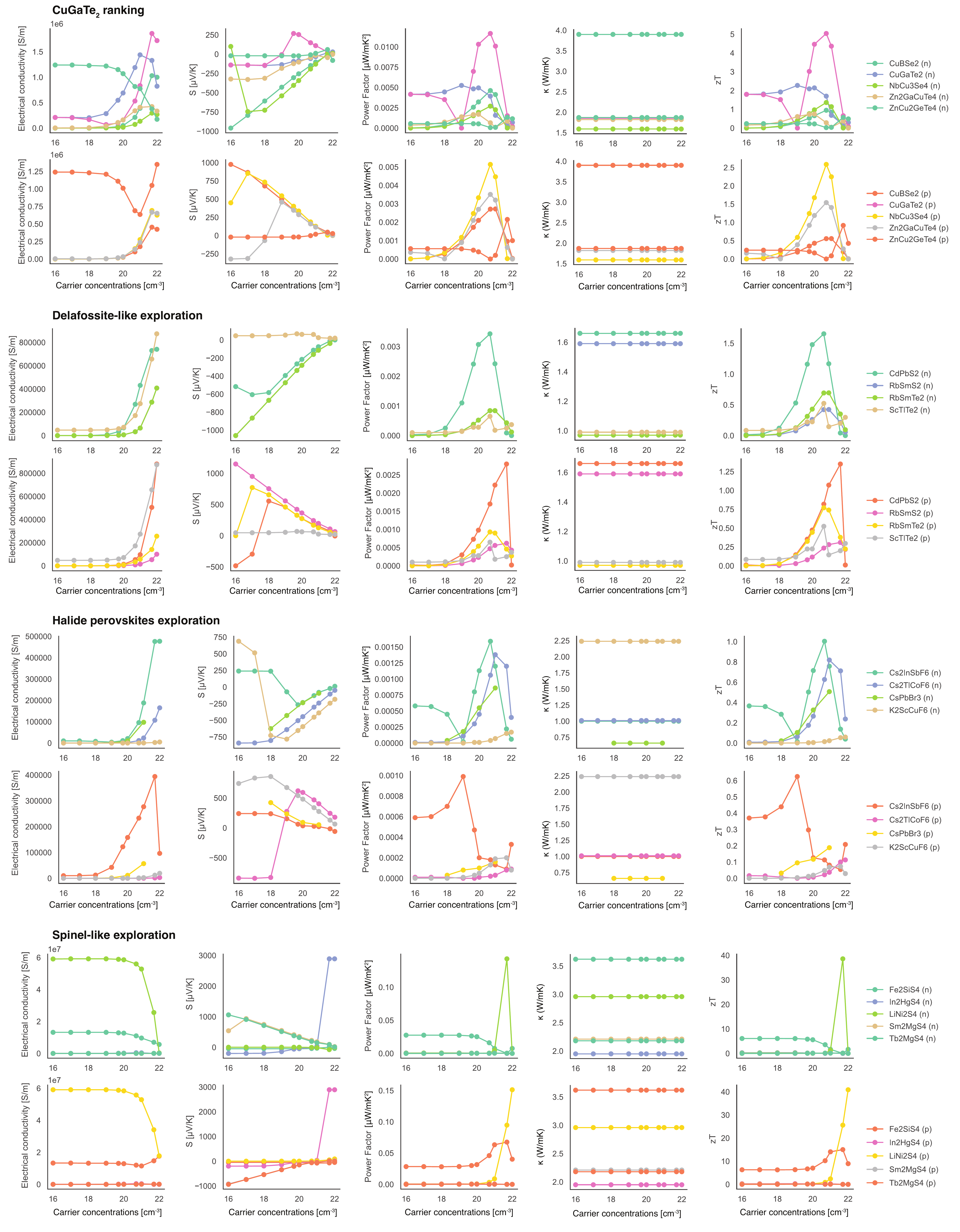} 
\caption{
First-principles calculations of TE properties for recommended materials from search ranking tasks and exploration tasks.}
\end{figure}

\newpage
\textbf{Evaluation: Experimental transport measurements on TE properties of CuZn$_2$GaTe$_4$}
\begin{figure}[!h]
\centering
\includegraphics[width=0.9\linewidth]{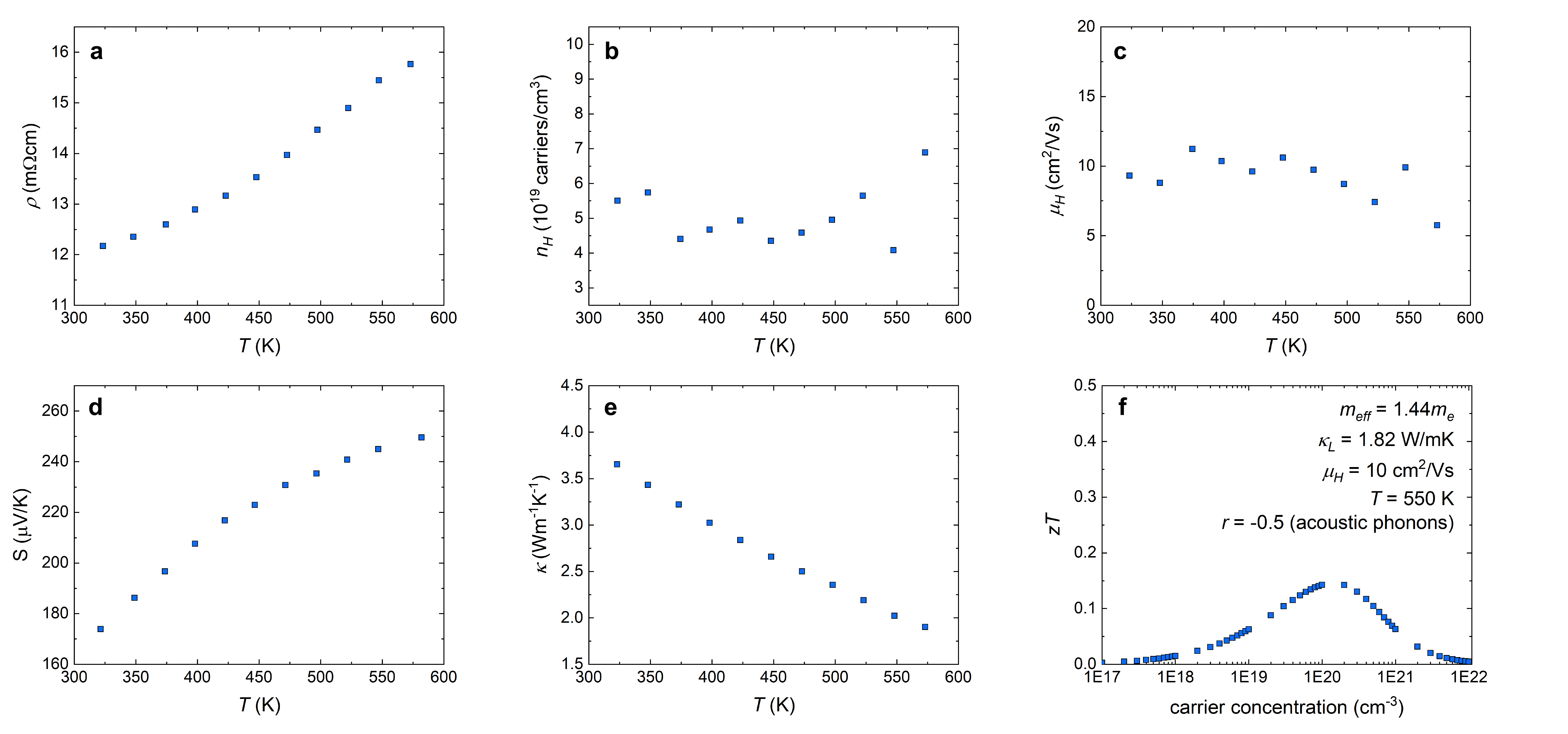}
\caption{
Experimental results on temperature-dependent TE properties of CuGaTe$_2$ `relevant' material CuZn$_2$GaTe$_4$. Panel display electrical resistivity (\textbf{a}), carrier concentration (\textbf{b}), carrier mobility (\textbf{c}), Seebeck coefficient (\textbf{d}), thermal conductivity (\textbf{e}), and figure of merit calculated with parabolic band model formalism using the experimental input parameters (\textbf{f}).}
\end{figure}

\textbf{Evaluation: Synthesis (experiment) on top candidates from ranking tasks}
\begin{figure}[!h]
\centering
\includegraphics[width=1\linewidth]{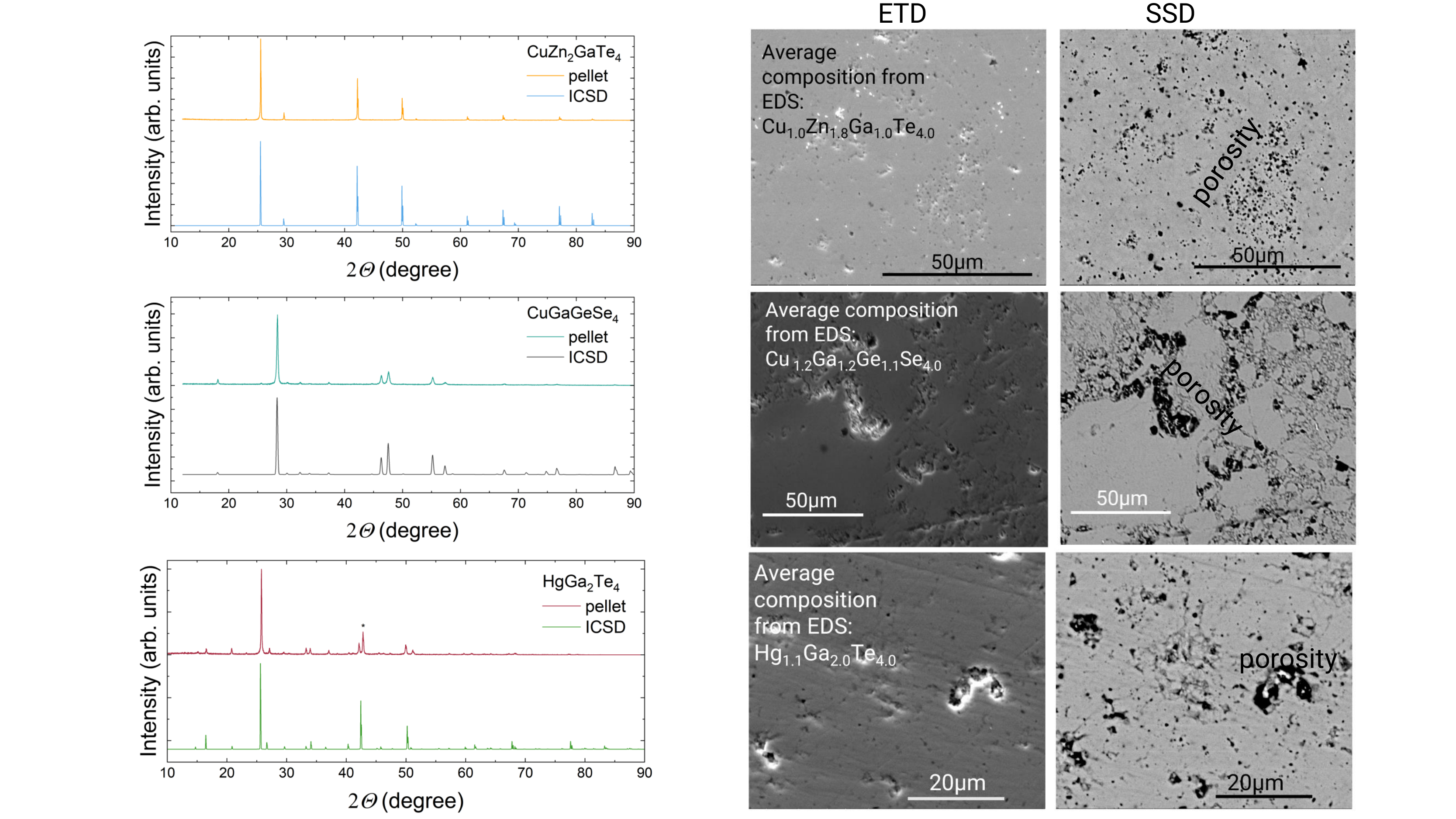}
\caption{All samples show porosity but excellent x-ray diffraction patterns. Left column of SEM images (ETD) corresponds to secondary electron emission, and right column (SSD) corresponds to imaging performed with back-scattered electrons. We observe good agreement between expected stoichiometry and stoichiometry measured from energy dispersive spectroscopy (EDS), which was performed over several points across the sample and averaged.
}
\end{figure}



\end{document}